\documentclass[12pt]{iopart}
\usepackage{amssymb}
\usepackage{graphicx}

\begin{document}

\title[Metrology and $1/f$ noise]{Metrology and $1/f$ noise: linear regressions and confidence intervals in flicker noise context}

\author{F. Vernotte$^1$ and E. Lantz$^2$}

\address{$^1$ UTINAM/Observatory THETA, University of Franche-Comt\'{e} and CNRS, 41 bis avenue de l'observatoire, BP 1615, 25010 Besan\c{c}on Cedex, France}
\address{$^2$ Department of Optics/Femto-ST, University of Franche-Comt\'{e} and CNRS, route des Montboucons, 25000 Besan\c{c}on, France}
\ead{francois.vernotte@obs-besancon.fr}
\vspace{10pt}

\begin{abstract}
$1/f$ noise is very common but is difficult to handle in a metrological way. After having recalled the main characteristics of a strongly correlated noise, this paper will determine relationships giving confidence intervals over the arithmetic mean and the linear drift parameters. A complete example of processing of an actual measurement sequence affected by $1/f$ noise will be given. 
\end{abstract}

%
\vspace{2pc}




\section{Introduction}
The flicker noise, or $1/f$ noise, may be encountered everywhere from atomic physics to astrophysics through  nano-technologies, electronics, \ldots\cite{press1978}.  Although its origin is better understood \cite{liu2013}, it remains a difficult issue and the nightmare of metrologists because of the strong correlations of its samples inducing a fundamentally duration dependent behavior. For example, unlike the white noise, the $1/f$ noise does not decrease by averaging but remains almost the same. Moreover, this is one of the way to be faced with the flicker noise: very often, we observe that the dispersion of measurements decreases as $1/\sqrt{N}$, where $N$ is the number of averaged measurements, until a certain value of $N$ for which the decrement stops. The flicker floor is reached.
It is then of importance to identify when we pass from a white noise to a $1/f$ context and what is the optimal average number.

However, once the flicker floor is reached, it is still possible to perform metrology but some precautions must be taken. Firstly, we must keep in mind that the $1/f$ noise takes its name from the dependency of its spectral density versus frequency: it means that the spectral density tends toward infinity for $f=0$. We have thus to ensure the convergence of the statistical parameters, such as the mean, by introducing a low cut-off frequency, below which the spectral density tends toward 0. But the existence of such a low cut-off frequency may be puzzling. This paper will give some clues to understand its physical meaning and the way to model it. Then, it is necessary to be able to define confidence intervals in such a context. Of course, the classical relationships which are designed for white noise are not valid for flicker.

This paper intends to determine rigorously new relationships giving confidence intervals over statistical parameters (arithmetic mean, drift coefficients) versus the number of measurements, the variance of the residuals and the hypothetic low cut-off frequency. In order to obtain such relationships, approximations will be performed on the autocorrelation function of the $1/f$ noise and on the variance calculation. These results will be validated by both numerical computations and Monte-Carlo simulations. Then, a methodology will be proposed for handling properly measurements in a $1/f$ context. Finally, this method will be applied to experimental cases.

But more than practical recipes, this paper aims to give a general method for finding such relationships for other types of noise in other contexts and for carefully validating the results.

\section{Problem statement}
	\subsection{Measurement principle}
\begin{figure}
\centering{\includegraphics[width=7.5cm]{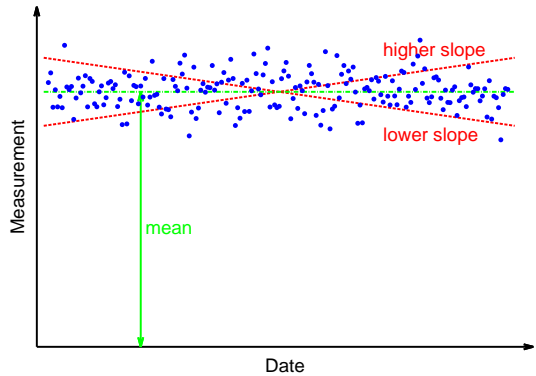}}
\caption{Measurement principle.\label{fig:princ}}
\end{figure}

Let us consider that we want to measure a quantity, e.g. a duration $D$, meant to be constant. In order to refine this measure and to verify the constancy of this quantity, we may perform several measurements, say $N$, at different dates and compute a linear regression over these measurements (see figure \ref{fig:princ}). Let us denote $d_i$ the measurements:
\begin{equation}
d_i= C_0 + C_1 t_i + \epsilon_i\label{eq:cl_reg_lin}
\end{equation}
where $C_0$ and $C_1$ are, respectively, the constant and the linear coefficients of the drift, $t_i$ the date of the measurement $d_i$ and $\epsilon_i$ the measurement noise, i.e. the random fluctuations of the measurements.

An estimation\footnote{Throughout this paper, we will distinguish a quantity $\theta$ from its estimate $\hat{\theta}$ by adding a hat $\hat{}$ at the top of the estimate symbol.} of the quantity $D$ may be obtained thanks to the arithmetical mean:
\begin{equation}
\hat{D}=\frac{1}{N} \sum_{i=0}^{N-1}d_i.
\end{equation}

The linear regression provides an estimation of the drift coefficients: $\hat{C}_0$ and $\hat{C}_1$.

We can then extract the residuals $e_i$ as the difference between the measurements and the estimated drift:
\begin{equation}
\e_i=d_i - \hat{C}_0 - \hat{C}_1 t_i,
\end{equation}
and compute the variance of the residuals $\sigma_e^2$.

Since the quantity $D$ is supposed to be constant, the slope of the drift should be null. We have then to verify that the estimate $\hat{C}_1$ is compatible with 0, i.e. that the uncertainty over the slope estimate is larger than the estimate:
\begin{equation}
\Delta C_1 > \hat{C}_1.
\end{equation}

We also have to define a confidence interval $\Delta D$ around the estimate $\hat{D}$:
\begin{equation}
\hat{D}-\Delta D < D < \hat{D}+\Delta D \quad \textrm{@ } 95 \textrm{ \% confidence.}
\end{equation}

The main issue is then: how is it possible to estimate the uncertainties $\Delta D$, $\Delta C_0$ and $\Delta C_1$ from the variance of the residuals $\sigma_e^2$? The answer is well known if the random fluctuations are following a Laplace-Gauss distribution, i.e. $\{\epsilon_i\}$ is a white Gaussian noise, but what happens if $\{\epsilon_i\}$ is a $1/f$ noise? That is the purpose of this paper.

	\subsection{White noise versus strongly correlated noises}
Let us consider that the $N$ measurements were regularly spaced and were performed with a sampling step $\tau_0$: $t_i=i\tau_0$.

The drift coefficients are computed from these relationships \cite{kenney1962}:
\begin{eqnarray}
\hat{C}_0 &=& \displaystyle\frac{2(2N+1)}{(N-1)N}\sum_{i=1}^N d_i + \frac{-6}{(N-1)N}\sum_{i=1}^N i d_i\\
\hat{C}_1 &=& \displaystyle\frac{-6}{(N-1)N\tau_0}\sum_{i=1}^N d_i + \frac{12}{(N-1)N(N+1)\tau_0}\sum_{i=1}^N i d_i
\end{eqnarray}

Since the residuals are centered, the variance of the residuals may be computed as:
\begin{equation}
\sigma_e^2=\frac{1}{N}\sum_{i=1}^N e_i^2.
\end{equation}

Obviously, the computation of the uncertainties depends on the distribution of the random fluctuations.

		\subsubsection{Case of a white noise.}
If $\{\epsilon_i\}$ is a white Gaussian noise, the variance of the drift coefficients are given by \cite{vernotte2001}:
\begin{eqnarray}
\sigma_{C0}^2=\frac{2(2N+1)}{N(N-1)} \sigma_e^2 &\approx & \frac{4}{N} \sigma_e^2 \qquad \textrm{(for large }N\textrm{)}\\
\sigma_{C1}^2=\frac{12}{N(N-1)(N+1)\tau_0^2} \sigma_e^2 &\approx & \frac{12}{N^3\tau_0^2} \sigma_e^2.
\end{eqnarray}

The 95 \% uncertainty domains over $C_0$ and $C_1$ may be assessed as $\Delta C_0=2\sigma_{C0}$ and  $\Delta C_1=2\sigma_{C1}$.

If the drift may be considered as null, i.e. $\hat{C}_1 < \Delta C_1$, $D$ may be assumed as constant and we can estimate a 95 \% confidence interval over $\hat{D}$: 
\begin{equation}
\Delta D = \frac{2}{\sqrt{N}}\sigma_e \qquad \textrm{(for large }N\textrm{)}
\end{equation}
(if $N<20$, the Student coefficients must be used \cite{saporta1990}).

Thus, in the case of a white noise, the uncertainty over $\hat{D}$ decreases as $1/\sqrt{N}$, it is then very useful to perform a huge number of measurements for reducing $\Delta D$.

		\subsubsection{Case of a strongly correlated noise.}
\begin{figure}
\centering{\includegraphics[width=7.5cm]{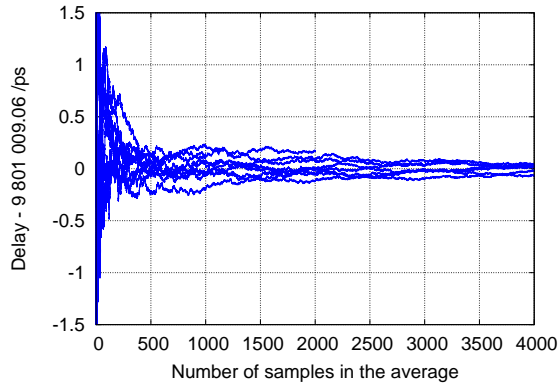}}
\caption{Cumulative average of delay measurements minus the arithmetic mean ($\hat{D}\approx 10$ $\mu$s). The dispersion of the average decreases up to 1\ 000 measurements and remains constant for a larger number of samples.\label{fig:cumu}}
\end{figure}

However, a limitation of the decreasing of the uncertainty versus the number of measurements is generally observed. Figure \ref{fig:cumu} presents the evolution of the arithmetic mean versus the number of samples in the case of delay measurements (see more details about this experiment in \cite{meyer2014}). In this example, it is particularly clear that it is useless to average more than 1000 samples.

\begin{figure}
\centering{\includegraphics[width=7.5cm]{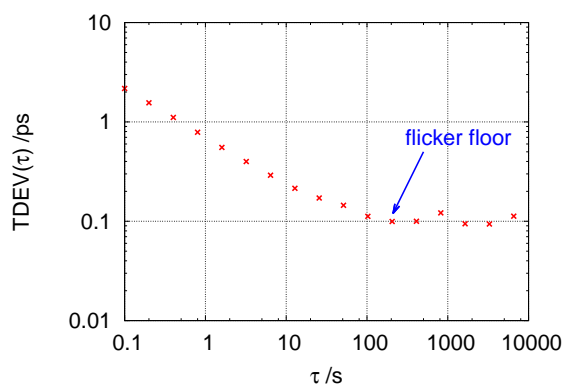}}
\caption{TDev curve of delay measurements (sampling rate: 10 Hz). The flicker floor is reached at 200 s (2\ 000 measurements).\label{fig:tvar}}
\end{figure}

In the time and frequency metrology domain, the Time Deviation estimator (TDev) \cite{allan1990, allan1991} is generally used to evaluate the limit number above which the average remains constant (the flicker floor). Figure \ref{fig:tvar} shows such a TDev curve in the case of the same experiment as figure \ref{fig:cumu}.

Once this limit is reached, the random fluctuations of the measurement are no longer a white Gaussian process and we have to deal with statistically duration dependent process:
\begin{itemize}
	\item the statistical parameters (mean, standard deviation, drift coefficients, \ldots) depend on when they are measured
	\item the statistical parameters depend on the duration of the measurement sequence
	\item the statistical parameters do not converge if this duration tends toward infinity!
\end{itemize}

Several types of statistically duration dependent noises exist (random walk, random run, \ldots) but this paper will focus on the flicker noise because it is generally the first type of strongly correlated noise which is encountered after the white noise limit.

\begin{figure}
\centering{\includegraphics[width=7.5cm]{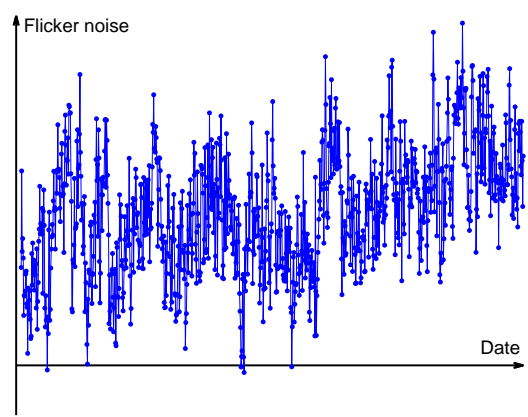}}
\caption{Example of flicker noise.\label{fig:flic}}
\end{figure}

Figure \ref{fig:flic} presents an example of realization of flicker noise. Depending on the region which is observed, the plot exhibits different behaviors with various means, dispersions or trends (the beginning of a ``false drift'' can even be seen at the end of the sequence). Moreover, the mean of the samples is clearly positive. We can bet that if we waited longer, the mean would depart more from the origin (above or below) and would tend toward infinity if the sequence length tend also toward infinity.

In other words, some statistical parameters (the mean in this case) diverge for long sequence unless a low cut-off frequency of the signal is introduced.

	\subsection{Low cut-off frequency, PSD and autocorrelation function}
But, what is the physical meaning of such a low cut-off frequency? Is it the inverse of the duration of the sequence, of the duration from which the measurement device is powered, of the age of this device or of the age of the Universe? Neither of them, the next sections will attempt to answer this question.

\subsubsection{The moment condition.\label{sec:moment_condition}}
In order to answer to that question, we need to use the ``moment condition'' \cite{deeter1982,vernotte2002}. Let us consider a linear estimator $h(t)$ which gives an estimate $\hat{\theta}$ of the quantity $\theta$ from $N$ measurements $\{d_i\}$:
\begin{equation}
\hat{\theta}=\sum_{i=0}^{N-1}h(t_i) d_i=\sum_{i=0}^{N-1}h_i d_i\label{eq:estpi}
\end{equation} 
where $\theta$ may be either the mean value $D$, the constant drift coefficient $C_0$, the linear drift coefficient $C_1$, \ldots

Assuming that the mathematical expectation of $\theta$ is null ($\theta$ may either be positive or negative), the variance of this estimate is then:
\begin{equation}
\sigma_\theta^2=\mathrm{E}(\theta^2)=\mathrm{E}\left\{\left[\sum_{i=0}^{N-1}h_i d_i\right]^2\right\}
\end{equation}
where the symbol $E(q)$ stands for the mathematical expectation of the quantity $q$.

Let us also define $H(f)$, the Fourier transform of the estimator $h(t)$. $H(f)$ is then the transfer function of the estimator (of the filter) $h(t)$. Assuming that the frequency samples $\{H(f_i)\}$ ($f_i$ $\in$ $\left\{0, 1/(N\tau_0), 2/(N\tau_0), \ldots, 1/(2\tau_0)\right\}$) are uncorrelated, the variance $\sigma_\theta^2$ may be written likewise as (see Appendix):
\begin{equation}
\sigma_\theta^2=\int_{-\infty}^{+\infty}\left|H(f)\right|^2 S_d(f) \mathrm{d}f\label{eq:est_f}
\end{equation}
where $S_d(f)$ is the power spectral density (PSD) of the measurements $\left\{d_i\right\}$ (see (\ref{eq:defPSD}) in Appendix for the mathematical definition of the PSD).

The moment condition establishes the equivalence between the sensitivity of the estimator $h(t)$ for drifts and its convergence for low frequency noises according to the following inequality:
\begin{equation}
\begin{array}{c}
\displaystyle\int_{-\infty}^{+\infty}\left|H(f)\right|^2 f^\alpha \mathrm{d}f \quad \textrm{converges}\\
\Leftrightarrow \quad \displaystyle\sum_{i=0}^{N-1} h_i t_i^q=0 \quad \textrm{for} \quad 0\leq q \leq \frac{-\alpha -1}{2}.
\end{array}
\end{equation}
In the case of a flicker noise, $\alpha=-1$ and the moment condition becomes:
\begin{equation}
\int_{-\infty}^{+\infty}\left|H(f)\right|^2 f^{-1} \mathrm{d}f \quad \textrm{converges} \quad \Leftrightarrow \quad \sum_{i=0}^{N-1} h_i t_i^0=\sum_{i=0}^{N-1} h_i = 0.
\end{equation}
What implies this condition on the three estimators that interest us: the arithmetic mean, the constant drift coefficient and linear drift coefficient?

The arithmetic mean may be assessed in this way by using $h_m(t)=1/N$. Therefore, $\sum_{i=0}^{N-1} h_{mi} \neq 0$ proving that it does not converge.

The estimator of the constant drift coefficient is designed for estimating a quantity which is directly linked to the mean of the measurement sequence. In other words, if such a mean is not null, this estimator will measure it. Therefore, it cannot give a null result and $\sum_{i=0}^{N-1} h_{mi} \neq 0$ proving that this estimator does not converge. An example of such an estimator, $\phi_0(t)=1/\sqrt{N}$, will be given in {\S} \ref{sec:cheb} ``Chebyshev polynomials''. It is clear that the estimator does not converge either.

On the other hand, estimators of the linear coefficient drift may be constructed in such a way that they are not sensitive to a constant (it is an advantage since it ensures the independence between the 2 drift coefficient estimations). Therefore, these estimators converge for a flicker noise.  An example of such an estimator, $\phi_1(t)$, see (\ref{eq:tcheby}), will be given in {\S} \ref{sec:cheb} ``Chebyshev polynomials''.

In other words, a flicker noise exhibits a mean value which does not converge, i.e. which increases infinitely if the duration of the sequence increases, whereas its ``natural drift'' (the false drift of figure \ref{fig:flic}) is independent on the duration of the sequence. However, we must keep in mind that the mathematical expectation of these statistical quantities (arithmetic mean, constant and linear drift coefficient) is equal to zero, only their variances are not null. To summarize, the variance of the linear drift coefficient does not depend on the low cut-off frequency (as in the case of an uncorrelated noise) but the mean and the constant drift coefficient depends on it and diverge if the low cut-off frequency tends toward zero.

As a consequence, the only visible effect of the low cut-off frequency is to increase the mean of the sequence.

	\subsubsection{Meaning of the low cut-off frequency.\label{sec:fl}}
The answer of the question of the beginning of this section, i.e. what is the low cut-off frequency, is now obvious: removing the mean value of a flicker sequence is equivalent to set to zero the amplitude of the spectrum at $f=0$ and, therefore, to set the low cut-off frequency to the inverse of the duration of the sequence $f_l=1/(N\tau_0)$! The meaning of the low cut-off frequency is then the inverse of the duration over which we subtract the arithmetic mean. We will see in {\S} \ref{sec:vsfl} that this definition of the low cut-off frequency needs to be developed but it gives interesting clues for understanding.

However, the metrological consequences of the removal of the mean value may be puzzling: the arithmetic mean of the residuals (after removing the mean value) is obviously identically null and therefore the variance of this mean is equal to 0! Would it mean that the estimation of the quantity $D$ is certain, i.e. that $\Delta D =0$? No, of course, it means that the mean value of the flicker noise accounts for the accuracy (or rather the inaccuracy) of the estimation of $D$ and its estimation over the next measurement sequence will be different. If we want a confidence interval over both measurement sequences, we have to consider a low cut-off frequency equal to the inverse of the total duration of these sequences, i.e. the date of the end of the second sequence minus the date of the beginning of the first sequence.

We need then to know the expression of the variance of our three statistical quantities versus the number of measurements $N$ and the low cut-off frequency $f_l$. In order to perform this calculation, we have to model the PSD of the flicker noise including its low cut-off frequency.

	\subsubsection{Modeling the Power Spectral Density of a flicker noise.}
\begin{figure}
\centering{\includegraphics[width=7.5cm]{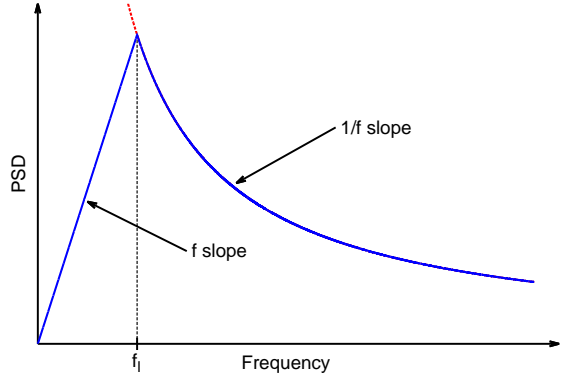}}
\caption{PSD model with a low frequency noise $f_l$.\label{fig:PSD}}
\end{figure}

A flicker noise is also called $1/f$ noise because its PSD decreases as the inverse of the frequency. But what is its behavior below the low cut-off frequency? We can assume that the low cut-off frequency is due to a ``natural'' first-order high-pass filter: below $f_l$ the high pass filter has a $f^2$ slope and above it is constant and equal to 0. Therefore, the PSD $S_d(f)$ increases as $f$ below $f_l$ and decreases as $1/f$ above $f_l$ without discontinuity\footnote{For the sake of simplicity, we did not model the high-pass transfer function by $f^2/(f+f_l)^2$.} (see figure \ref{fig:PSD}): 
\begin{equation}
\left\{\begin{array}{lll}
S_d(f) = k_{-1} f/f_l^2 & \quad & \textrm{for} \quad f<f_l\\
S_d(f) = k_{-1}/f & \quad & \textrm{for} \quad f_l<f<f_h\\
S_d(f) = 0 & \quad & \textrm{for} \quad f>f_h
\end{array}\right.\label{eq:PSD}
\end{equation}
where $k_{-1}$ is the flicker noise level. The high cut-off frequency $f_h$, which is also necessary to ensure convergence to the high frequencies, is provided by the sampling process: $f_h=1/(2\tau_0)$. 

	\subsubsection{Autocorrelation function.} 
The autocorrelation function $R_d(\tau)$ is the Fourier transform of the PSD $S_d(f)$. It expresses the correlation between two samples separated by a duration $\tau$ : $R_d(\tau)=\mathrm{E}\left[d(t)d(t+\tau)\right]$. 

We can then calculate the autocorrelation function from the PSD:
\begin{equation}
R_d(\tau)=\int_{-\infty}^{+\infty} S_d^{TS}(f) e^{j2\pi \tau f}\mathrm{d}f\label{eq:def_R}
\end{equation}
where 
 $S_d^{TS}(f)$ represents the ``two sided'' PSD, i.e. defined over $\mathbb{R}$. Since $S_d^{TS}(f)$ is even, we prefer generally used the ``one sided'' PSD defined over $\mathbb{R}^+$:
\begin{equation}
\left\{\begin{array}{lll}
S_d(f) = 2 S_d^{TS}(f) & \quad & \textrm{for} \quad f\geq 0\\
S_d(f) = 0 & \quad & \textrm{for} \quad f<0.
\end{array}\right.\label{eq:STS}
\end{equation}
From relationships (\ref{eq:def_R}) and (\ref{eq:STS}) and because $S_d(f)$ is purely real, the autocorrelation function can be rewritten as:
\begin{equation}
R_d(\tau)=\int_{0}^{+\infty} S_d(f) \cos(2\pi\tau f) \mathrm{d}f.
\end{equation}
Replacing $S_d(f)$ by its model given in (\ref{eq:PSD}), it comes:
\begin{equation}
R_d(\tau)=k_{-1} \left[\int_{0}^{f_l} \frac{f}{f_l^2} \cos(2\pi\tau f) \mathrm{d}f + \int_{f_l}^{f_h} \frac{\cos(2\pi\tau f)}{f} \mathrm{d}f\right].\label{eq:Rd}
\end{equation}


The resolution of this integral gives:
\begin{equation}
\left\{\begin{array}{rcl}
R_d(0)&=&k_{-1}\left[\frac{1}{2}+\ln(fh/fl)\right]\\
R_d(\tau)&=&\displaystyle k_{-1}\left[\frac{\cos(2\pi f_l\tau)-1+2\pi f_l\tau\sin(2\pi f_l\tau)}{(2\pi f_l\tau)^2}\right.\\
&&\left.\vphantom{\displaystyle\frac{\cos(2\pi f_l\tau)-1+2\pi f_l\tau\sin(2\pi f_l\tau)}{(2\pi f_l\tau)^2}}+\mathrm{Ci}(2\pi\tau f_h)-\mathrm{Ci}(2 \pi \tau f_l)\right]
\end{array}\right.\label{eq:exRx}
\end{equation}
where the Cosine Integral function $\mathrm{Ci}(x)$ is defined as \cite{abramowitz1972}:
\begin{equation}
\forall x > 0,\ \mathrm{Ci}(x) = -\int_{x}^\infty \frac{\cos(y)}{y} \mathrm{d}y.
\end{equation}

Let us consider that the low cut-off frequency is very low, i.e. $f_l \ll 1/(N\tau_0)$. This condition is not restrictive since we know that the effect of $f_l$ is limited to the mean value of the sequence.  Therefore, we can consider in the first term of (\ref{eq:Rd}) that $2\pi\tau f \leq 2\pi N\tau_0 f_l \ll 2\pi$ and then that $\cos(2\pi\tau f)=1$. Thus, (\ref{eq:exRx}) may be rewritten as:
\begin{equation}
\left\{\begin{array}{l}
R_d(0)=k_{-1}\left[\frac{1}{2}+\ln(fh/fl)\right]\\
R_d(\tau)=k_{-1}\left[\frac{1}{2}+\mathrm{Ci}(2\pi\tau f_h)-\mathrm{Ci}(2 \pi \tau f_l)\right]
\end{array}\right.\label{eq:spRx}
\end{equation}

The Taylor expansion of $\mathrm{Ci}(x)$, in the neighborhood of 0, is given by \cite{abramowitz1972}:
\begin{equation}
\forall x \in \mathbb{R}^{+*},\ \mathrm{Ci}(x) = C + \ln(x) + \sum_{n=1}^{+\infty} (-1)^n \frac{x^{2n}}{(2n)!(2n)} \label{eq:dlCI}
\end{equation}
where $C\approx 0.5772$ is the Euler-Mascheroni constant.

Since $\tau \in \left\{\tau_0, 2\tau_0, \ldots, N\tau_0\right\}$, $\mathrm{Ci}(2\pi\tau f_h)=\mathrm{Ci}(k \pi)$ with $k \in \left\{1, 2, \ldots, N\right\}$, thus $\mathrm{Ci}(2\pi\tau f_h)$ will oscillate around 0. On the other hand, $2 \pi \tau f_l \ll 1$ and $-\mathrm{Ci}(2 \pi \tau f_l)$ may be approximated by its Taylor expansion at first order: $-\mathrm{Ci}(2 \pi \tau f_l)\approx -C - \ln(2 \pi \tau f_l)$. Therefore, $\mathrm{Ci}(2\pi\tau f_h)$ is negligible regarding $-\mathrm{Ci}(2 \pi \tau f_l)$. 

This leads to an approximated expression of $R_d(\tau)$ for $\tau\neq 0$:
\begin{equation}
R_d(\tau) \approx k_{-1} \left[\frac{1}{2} -C - \ln|2 \pi \tau f_l|\right].\label{eq:apRx}
\end{equation}

It may be noticed that this expression differs from the one previously published in \cite{vernotte2001} because the $f$ slope behavior of $S_d(f)$ below $f_l$ was not taken into account. 

\begin{figure}
\centering{\includegraphics[width=7.5cm]{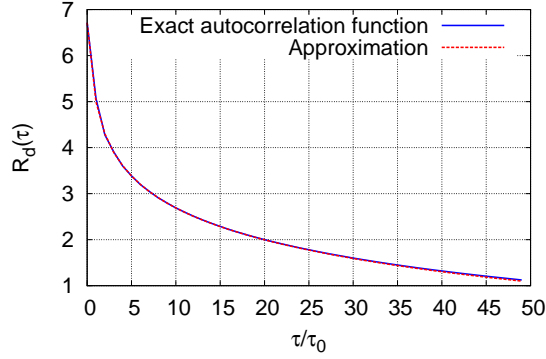}}
\caption{Comparison of the exact correlation function and the approximation of (\ref{eq:apRx}) for $N=50$ data and $f_l=1/(20 N\tau_0)$. The larger difference between these expressions is equal to 2.1 \%.\label{fig:Rx}}
\end{figure}

Figure \ref{fig:Rx} compares the exact expression of $R_d(\tau)$ from relationship (\ref{eq:exRx}) with the approximation given in (\ref{eq:apRx}) in the case of $N=50$ measurements and the inverse of the low cut-off frequency 20 times larger than the duration of the sequence. This graph shows that the approximation (\ref{eq:apRx}) is perfectly valid. In the following, we will then consider that the autocorrelation of a flicker noise is given by:
\begin{equation}
\left\{\begin{array}{l}
R_d(0)=\displaystyle k_{-1}\left[\frac{1}{2}+\ln(fh/fl)\right]\\
R_d(\tau)\approx\displaystyle k_{-1} \left[\frac{1}{2} -C - \ln|2 \pi \tau f_l|\right].
\end{array}\right.\label{eq:approx_Rd}
\end{equation}

We now ought to calculate the variance of the drift coefficients $C_0$ and $C_1$. However, this task is not so simple because these parameters are not statistically optimized: they depends on the sampling and, above all, they are strongly correlated. The problem will be far easier if we adopt 
 a linear fit by using the Chebyshev polynomials.

	\subsection{Estimation with Chebyshev polynomials\label{sec:cheb}}
Rather than the classical linear regression of equation (\ref{eq:cl_reg_lin}), let us use the first two Chebyshev polynomials $\Phi_0(t)$ and $\Phi_1(t)$, i.e. the Chebyshev polynomials of degrees, respectively, 0 and 1:
\begin{equation}
d_i= P_0 \Phi_0(t_i) + P_1 \Phi_1(t_i)+\epsilon_i\label{eq:Cheb_reg_lin}
\end{equation}
with
\begin{equation}
\left\{\begin{array}{l}
\Phi_0(t) = \displaystyle \frac{1}{\sqrt{N}} \\
\Phi_1(t) = \displaystyle \sqrt{\frac{3}{(N-1)N(N+1)}}\left[2\frac{t}{\tau_0}-(N-1)\right].
\end{array}\right.\label{eq:tcheby}
\end{equation}

\begin{figure}
\centering{\includegraphics[width=7.5cm]{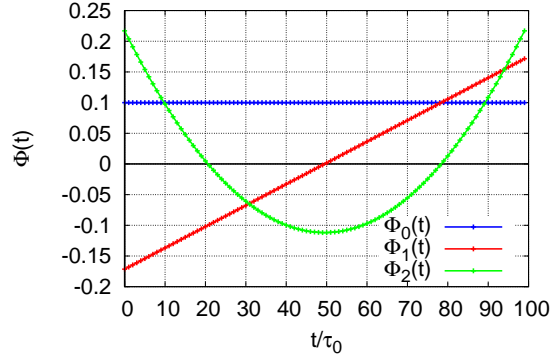}}
\caption{The first 3 Chebyshev polynomials for $N=100$.\label{fig:tcheby}}
\end{figure}

	\subsubsection{Properties of the Chebyshev polynomials.}
The main advantage of this approach lies in the orthonormality of these polynomials:
\begin{equation}
\sum_{i=0}^{N-1}\Phi_j(t_i)\Phi_k(t_i)=\delta_{ij} \quad \textrm{with } i \textrm{ and } j \in \{0,1\}\label{eq:orthon}
\end{equation}
where $\delta_{jk}$ is the Kronecker's delta. Therefore, the different Chebyshev polynomials are uncorrelated and normalized. Moreover, they are dimensionless and the dimension of the problem, e.g. the time in our example, is supported by the coefficients $P_0$ and $P_1$.

	\subsubsection{Calculation of the drift coefficient estimates.}
We ought to search the estimates $\hat{P}_0$ and $\hat{P}_1$ which will minimize the residuals $\{e_i\}$:
\begin{equation}
d_i=\hat{P}_0 \Phi_0(t_i)+\hat{P}_1 \Phi_1(t_i)+e_i.\label{eq:Cheb_est}
\end{equation} 

How can we obtained these estimates? Let us calculate $\mathrm{E}\left[\sum_{i=0}^{N-1}\Phi_0(t_i) d_i\right]$ and $\mathrm{E}\left[\sum_{i=0}^{N-1}\Phi_1(t_i) d_i\right]$. From (\ref{eq:Cheb_est}), it comes:
\begin{equation}
\left\{\begin{array}{rcl}
\displaystyle \mathrm{E}\left[\sum_{i=0}^{N-1}\Phi_0(t_i) d_i\right] &=& \displaystyle \mathrm{E}(\hat{P}_0) \sum_{i=0}^{N-1} \Phi_0^2(t_i)\\ 
&&\displaystyle + \mathrm{E}(\hat{P}_1)\sum_{i=0}^{N-1} \Phi_0(t_i)\Phi_1(t_i)\\
&&\displaystyle +\mathrm{E}\left[\sum_{i=0}^{N-1}\Phi_0(t_i)e_i\right] \\
\displaystyle \mathrm{E}\left[\sum_{i=0}^{N-1}\Phi_1(t_i) d_i\right] &=& \displaystyle \mathrm{E}(\hat{P}_0)\sum_{i=0}^{N-1} \Phi_0(t_i)\Phi_1(t_i) \\
&&\displaystyle + \mathrm{E}(\hat{P}_1) \sum_{i=0}^{N-1} \Phi_1^2(t_i)\\
&&\displaystyle +\mathrm{E}\left[\sum_{i=0}^{N-1}\Phi_1(t_i)e_i\right].
\end{array}\right.\label{eq:estP0P1}
\end{equation}
From (\ref{eq:orthon}), we know that $\sum_{i=0}^{N-1} \Phi_0^2(t_i)=\sum_{i=0}^{N-1} \Phi_1^2(t_i)=1$ and that $\sum_{i=0}^{N-1} \Phi_0(t_i)\Phi_1(t_i)=0$. Furthermore, since the residuals $\{e_i\}$ are, by construction, orthogonal to the interpolating function $\Phi_0(t)$ and $\Phi_1(t)$, (\ref{eq:estP0P1}) may be written as:
\begin{equation}
\left\{\begin{array}{rcl}
\displaystyle \mathrm{E}\left[\sum_{i=0}^{N-1}\Phi_0(t_i) d_i\right] &=& \mathrm{E}(\hat{P}_0)\\
\displaystyle \mathrm{E}\left[\sum_{i=0}^{N-1}\Phi_1(t_i) d_i\right] &=& \mathrm{E}(\hat{P}_1).
\end{array}\right.\label{eq:princ_estP0P1}
\end{equation}
Thus, the estimation of the $P_k$ coefficients ($k\in\{0,1\}$) is quite simple:
\begin{equation}
\hat{P}_k=\sum_{i=0}^{N-1}\Phi_k(t_i) d_i.\label{eq:estPk}
\end{equation}

	\subsubsection{From $P_0, P_1$ to $C_0,C_1$.}
At last, it is very easy to come back to the $C_0$ and $C_1$ coefficients of the classical linear regression by using the following inverse transform which is deduced from (\ref{eq:Cheb_reg_lin}) and (\ref{eq:tcheby}):
\begin{equation}
\left\{\begin{array}{rcl}
C_0 & = & \displaystyle \frac{1}{\sqrt{N}} P_0 - \sqrt{\frac{3(N-1)}{N(N+1)}} P_1\\
C_1 & = & \displaystyle \frac{2}{\tau_0}\sqrt{\frac{3}{(N-1)N(N+1)}} P_1.
\end{array}\right.\label{eq:P0P1toC0C1}
\end{equation}

Once obtained the variance of $P_0$ and $P_1$, we will use this inverse transform for deducing the variance of $C_0$ and $C_1$.

For more details about the Chebyshev polynomials see \cite{rivlin1974} and about their use for estimation see \cite{deeter1982} and \cite{vernotte2001}.

	\subsection{Calculation principle}
		\subsubsection{Estimation of the coefficient variances.}
Let us assume that a $\{d_i\}$ sequence is zero mean and without drift: $P_0=0$ and $P_1=0$. The estimates $\hat{P}_0$ and $\hat{P}_1$ calculated over this sequence will have the following properties:
\begin{equation}
\left\{\begin{array}{l}
\mathrm{E}(\hat{P}_k)=0\\
\mathrm{E}(\hat{P}_k^2)=\sigma_{Pk}^2
\end{array}\right. \quad \textrm{with } k \in \{0,1\}
\end{equation}
where $\sigma_{Pk}^2$ is the variance of the estimate $\hat{P}_k$.

This last equation provides the way for estimating the variances of the coefficients $P_0$ and $P_1$:
\begin{equation}
\sigma_{Pk}^2=\mathrm{E}(\hat{P}_k^2)=\sum_{i=0}^{N-1}\sum_{j=0}^{N-1}\Phi_k(t_i)\Phi_k(t_j)R_d(t_i-t_j) \quad \textrm{with } k \in \{0,1\}.\label{eq:varP0P1}
\end{equation}

		\subsubsection{Estimation of the variance of the residuals.} Since the residuals are centered, their variance is equal to their second raw moment: 
\begin{equation}
\sigma_e^2=\mathrm{E}\left(\frac{1}{N}\sum_{i=0}^{N-1}e_i^2\right).
\end{equation} 
From (\ref{eq:Cheb_est}), it comes:
\begin{eqnarray}
\sigma_e^2&=&\mathrm{E}\left\{\frac{1}{N}\sum_{i=0}^{N-1}e_i\left[d_i-\hat{P}_0\Phi_0(t_i)-\hat{P}_1\Phi_1(t_i)\right]\right\}\\
&=&\frac{1}{N}\left[\mathrm{E}\left(\sum_{i=0}^{N-1}e_i d_i\right)-\mathrm{E}\left(\hat{P}_0\sum_{i=0}^{N-1}e_i \Phi_0(t_i)\right)-\mathrm{E}\left(\hat{P}_1\sum_{i=0}^{N-1}e_i \Phi_1(t_i)\right)\right].
\end{eqnarray}
Since the residuals $\{e_i\}$ are orthogonal to $\Phi_0(t)$ and $\Phi_1(t)$, it comes:
\begin{eqnarray}
\sigma_e^2&=&\frac{1}{N}\mathrm{E}\left(\sum_{i=0}^{N-1}e_i d_i\right)\\
&=&\frac{1}{N}\left\{\mathrm{E}\left[\sum_{i=0}^{N-1}d_i^2-\hat{P}_0\sum_{i=0}^{N-1}\Phi_0(t_i)d_i-\hat{P}_1\sum_{i=0}^{N-1}\Phi_1(t_i)d_i\right]\right\}\\
&=&\frac{1}{N}\left[\sum_{i=0}^{N-1}R_d(0)-\mathrm{E}\left(\hat{P}^2_0\right)-\mathrm{E}\left(\hat{P}^2_1\right)\right].
\end{eqnarray}
At last, we obtain the following relationship:
\begin{equation}
\sigma_e^2=R_d(0)-\frac{1}{N}(\sigma_{P0}^2+\sigma_{P1}^2).\label{eq:var_res}
\end{equation}

\section{Calculations for the flicker noise}
Let us apply the general relationships (\ref{eq:varP0P1}) and (\ref{eq:var_res}), providing respectively the coefficient variances and the variance of the residuals, to the case of the flicker noise.

Since we know that the low cut-off frequency only impacts the mean of the sequence, let us assume that it is far lower than the inverse of the sequence duration: $f_l\ll1/(N\tau0)$. In this condition, the autocorrelation is given by (\ref{eq:exRx}) and (\ref{eq:apRx}).

In the following, we will only detail the calculation of  $\sigma_{P0}^2$. The Mathematica notebook detailing the calculations of  $\sigma_{P1}^2$ and $\sigma^2_e$ is available on request by sending an email to the corresponding author.
	\subsection{Variance of $P_0$}
The interpolating function $\Phi_0(t)$ is constant and equal to $1/\sqrt{N}$. Thus, (\ref{eq:varP0P1}) becomes the following sum:
\begin{equation}
\sigma_{P0}^2=\frac{1}{N}\sum_{i=0}^{N-1}\sum_{j=0}^{N-1}R_d[(i-j)\tau_0].\label{eq:varP0fm1}
\end{equation}
In order to separate the constant term and the term depending on $\tau$ in $R_d(\tau)$, (\ref{eq:apRx}) may be decomposed as:
\begin{equation}
R_d[(i-j)\tau_0]=k_{-1} \left[\frac{1}{2} -C - \ln(2 \pi \tau_0 f_l)\right] -k_{-1}\ln|i-j|.
\end{equation}

Since the autocorrelation function is even:
\begin{eqnarray}
\sigma_{P0}^2&=&\frac{k_{-1}}{N}\left\{\sum_{i=0}^{N-1}R_d(0)+2\sum_{i=1}^{N-1}\sum_{j=0}^{i-1} R_d[(i-j)\tau_0]\right\}\\
&=&k_{-1}\left\{\frac{1}{2}+\ln(f_h/f_l)\right.\nonumber\\
&&+(N-1)\left[\frac{1}{2} -C - \ln(2 \pi \tau_0 f_l)\right]\nonumber\\
&&\left.-\frac{2}{N}\sum_{i=1}^{N-1}\sum_{j=0}^{i-1} \ln|i-j|\right\}.\label{eq:SP0tot}
\end{eqnarray}

The last term may be approximated by an integral:
\begin{equation}
\sum_{i=1}^{N-1}\sum_{j=0}^{i-1} \ln|i-j|\approx \int_{1}^{N-1}\int_{0}^{x_i-1}\ln|x_i-x_j|\mathrm{d}x_j \mathrm{d}x_i
\end{equation}
The result of this integral is:
\begin{eqnarray}
\int_{1}^{N-1}\int_{0}^{x_i-1}\ln|x_i-x_j|\mathrm{d}x_j \mathrm{d}x_i&=&-\frac{3N^2}{4}+5N-2+\frac{(N-1)^2}{2}\ln(N-1)\nonumber\\
&\approx&\frac{N^2}{4}\left[-3+2\ln(N)\right] \quad \textrm{for large $N$}\label{eq:approx_bigN}
\end{eqnarray}

Assuming that $N$ is large, we may approximate (\ref{eq:SP0tot}) by:
\begin{equation}
\sigma_{P0}^2=\left[2-C-\ln\left(2\pi f_l N \tau_0\right)\right]N k_{-1}.\label{eq:var_P0_fm1}
\end{equation}

		\subsection{Variance of $P_1$}
\begin{equation}
\sigma_{P1}^2=\frac{3N}{4} k_{-1}.\label{eq:var_P1_fm1}
\end{equation}

		\subsection{Variance of the residuals}
\begin{equation}
\sigma_e^2=\left[-\frac{9}{4}+C+\ln\left(2\pi f_h N \tau_0\right)\right] k_{-1}.\label{eq:var_res_fm1}
\end{equation}

		\subsection{Drift coefficient variance versus the variance of the residuals}
Since the variance of the residuals $\sigma_e^2$ is more accessible than the flicker noise level $k_{-1}$, it is useful to express the variance of the drift coefficients versus $\sigma_e^2$. Thus, by assuming that $f_h=1/(2\tau_0)$ and from (\ref{eq:var_P0_fm1}), (\ref{eq:var_P1_fm1}) and (\ref{eq:var_res_fm1}), we get:
\begin{equation}
\left\{\begin{array}{l}
\sigma_{P0}^2=\displaystyle\frac{2-C-\ln\left(2\pi\right)-\ln\left(f_l N\tau_0\right)}{-\frac{9}{4}+C+\ln(\pi) +\ln(N)}N\sigma_e^2\\
\sigma_{P1}^2=\displaystyle\frac{3N\sigma_e^2}{-9+4C+4\ln(\pi) +4\ln(N)}.
\end{array}\right.
\end{equation}
These expressions may be simplified by replacing the constants by their numerical values:
\begin{equation}
\left\{\begin{array}{l}
\sigma_{P0}^2\displaystyle\approx \frac{\left[-0.4151-\ln\left(f_l N\tau_0\right)\right] N \sigma_e^2}{-2,112 + 4 \ln(N)}\\
\sigma_{P1}^2\displaystyle\approx \frac{3 N \sigma_e^2}{-2,112 + 4 \ln(N)}.
\end{array}\right.\label{eq:varP0P1Rs}
\end{equation}

We must keep in mind that these approximations are valid for large $N$ ($N\gtrapprox 20$). On the other hand, they were calculated by assuming that $f_l\ll1/(N\tau_0)$ but are also valid for larger $f_l$, i.e. up to $f_l\lessapprox 1/(4N\tau_0)$. In the case of $f_l=1/(N\tau_0)$, remember than $\sigma_{P0}^2=0$ (see {\S} \ref{sec:fl}).

Furthermore, these results are suboptimal since the least squares method (classical or with the Chebyshev polynomials) is optimized for uncorrelated noise. Since we are dealing with flicker noise, the samples are obviously correlated. In this case, the optimal solution is given by the Generalized Least Squares.

	\subsection{Another approach: the Generalized Least Squares (GLS)}
		\subsubsection{Matrix formalization of the problem.}
Let us define the following vectors:
\begin{itemize}
	\item $\vec{d}$ is the $N$-lines vector of the $d_i$ measurements
	\item $\vec{\epsilon}$ is the $N$-lines vector of the measurement noise $\epsilon_i$
	\item $\vec{P}$ is the 2-lines vector of the Chebyshev parameters $P_0$ and $P_1$.
\end{itemize}

We can also define the $N$-lines $\times$ 2-columns matrix $[\Phi]$ as the interpolating function matrix:
\begin{equation}
[\Phi]=\left(\begin{array}{cc}
\Phi_0(t_0) & \Phi_1(t_0) \\
\vdots & \vdots \\
\Phi_0(t_{N-1}) & \Phi_1(t_{N-1})
\end{array}\right).
\end{equation}

With these notations, (\ref{eq:Cheb_reg_lin}) becomes:
\begin{equation}
\vec{d}=[\Phi]\vec{P}+\vec{\epsilon}.\label{eq:matest}
\end{equation}

Multiplying by the transposed matrix $[\Phi]^T$, it comes:
\begin{equation}
[\Phi]^T\vec{d}=[\Phi]^T[\Phi]\vec{P}+[\Phi]^T\vec{\epsilon}.
\end{equation}
From (\ref{eq:orthon}) and because $\mathrm{E}\left([\Phi]^T\vec{\epsilon}\right)=0$, we obtain $\vec{\hat{P}}$, the 2-lines vector of the unbiased Chebyshev parameter estimates:
\begin{equation}
\vec{\hat{P}}=[\Phi]^T\vec{d}.\label{eq:optiwhite}
\end{equation}
It may be shown that $\vec{\hat{P}}$ of (\ref{eq:optiwhite}) is an optimal estimator of $\vec{P}$ if $\vec{\epsilon}$ is a white Gaussian noise, i.e. the $\{\epsilon_i\}$ are uncorrelated. 

The GLS, introduced by Aitken in 1934 \cite{aitken1934}, is a generalization of the least squares to correlated noises (and to measurements with unequal dispersions). It gives the optimal solution to (\ref{eq:matest}) when $\vec{\epsilon}$ is not a white Gaussian noise.

	\subsubsection{Optimal solution for a flicker noise.}
Let us define the noise covariance matrix $[C_\epsilon]=\mathrm{E}\left(\vec{\epsilon}\vec{\epsilon}^T\right)$:
\begin{equation}
[C_\epsilon]=\left(\begin{array}{cccc}
R_d(0) & R_d(\tau_0) & \ldots & R_d\left((N-1)\tau_0\right) \\
R_d(\tau_0) & R_d(0) & \ldots & R_d\left((N-2)\tau_0\right) \\
\vdots & \vdots & \ddots & \vdots \\
R_d\left((N-1)\tau_0\right) & R_d\left((N-2)\tau_0\right) & \ldots & R_d(0)
\end{array}\right).\label{eq:cov_Rd}
\end{equation}
From \cite{jenkins1968} (see equation (A4.1.4) in {\S} A4.1), the optimal solution $\vec{\hat{P}}^\star$ of (\ref{eq:matest}) is given by:
\begin{equation}
\vec{\hat{P}}^\star=[\Xi] [\Phi]^T [C_\epsilon]^{-1} \vec{d}\label{eq:opt_GLS}
\end{equation}
where the matrix $[\Xi]$ is defined as:
\begin{equation}
[\Xi]=\left([\Phi]^T[C_\epsilon]^{-1}[\Phi]\right)^{-1}.\label{eq:Theta}
\end{equation}

In the case of a white Gaussian noise, $[C_\epsilon]=[I_N]$, the unit matrix $N\times N$ and then $[\Xi]=\left([\Phi]^T[\Phi]\right)^{-1}$. Since the Chebyshev polynomials are orthonormal, $[\Phi]^T[\Phi]=[I_2]$ and $[\Xi]=[I_2]$. Therefore, (\ref{eq:opt_GLS}) reduces to (\ref{eq:optiwhite}).

In the case of a flicker noise, we have to use the expression of $R_d(\tau)$ given by (\ref{eq:spRx}) in (\ref{eq:cov_Rd}) and compute $\vec{\hat{P}}^\star$ from (\ref{eq:Theta}) and (\ref{eq:opt_GLS}).

Moreover, $[\Xi]$ is the covariance matrix of the $\vec{\hat{P}}^\star$ estimate vector and $[C_\epsilon]-[\Phi][\Xi][\Phi]^T$ is the covariance matrix of the residuals \cite{jenkins1968}. Hence:
\begin{equation}
\left\{\begin{array}{l}
\sigma_{P0\star}^2=\Xi_{1,1}\\
\sigma_{P1\star}^2=\Xi_{2,2}
\end{array}\right.\label{eq:varGLS}
\end{equation}
and
\begin{equation}
\sigma_{e\star}^2=\frac{1}{N}\Tr\left\{[C_\epsilon]-[\Phi][\Xi][\Phi]^T\right\}.\label{eq:resGLS}
\end{equation}

In {\S} \ref{sec:cmp_opt}, we will compare the GLS estimates (\ref{eq:opt_GLS}), their variances (\ref{eq:varGLS}) and the variance of the residuals (\ref{eq:resGLS}) to the ones given by the Chebychev Least Squares in (\ref{eq:estPk}), (\ref{eq:var_P0_fm1}), (\ref{eq:var_P1_fm1}) and (\ref{eq:var_res_fm1}).

\section{Validation of the theoretical results}
In order to validate these theoretical relationships, we performed two types of checking:
\begin{itemize}
	\item a comparison with numerical computations of (\ref{eq:varP0P1})
	\item a comparison with Monte-Carlo simulations obtained with a noise simulator.
\end{itemize} 
In all cases, we considered flicker noises with a unity level, i.e. $k_{-1}=1$.

	\subsection{Comparison with numerical computations}
We computed the result of equation (\ref{eq:varP0P1}) with the exact expression of the autocorrelation function given by (\ref{eq:exRx}) (the $\textrm{Ci}(x)$ function is available in the GNU Scientific Library as well as in Matlab and Octave). These computations were performed according to two protocols:
\begin{itemize}
	\item by varying $N$ with $f_l$ constant; we chose $N \in \{16, 64, 256, 1024, 4096, 16384\}$ data and $fl=1/(65536 \tau_0)$
	\item by varying $f_l$ with $N$ constant; we chose $\tau_0/f_l \in \{256, 512, 1024, 4096, 16384, 65536\}$ and $N=256$.
\end{itemize}
The results are plotted in figure \ref{fig:P01rsvsNfl} (green circles) and compared to the theoretical values (blue curves) given by relationships (\ref{eq:var_P0_fm1}), (\ref{eq:var_P1_fm1}) and (\ref{eq:var_res_fm1}).

	\subsection{Comparison with Monte-Carlo simulations}
The noise simulator we used requires the following input parameters: the inverse of the low cut-off frequency in terms of sampling step $1/f_l=M\tau_0$, the number of data $N$ (with $N\leq M$), the type of noise $\alpha$ and the noise level $k_\alpha$. It computes a $M$-sample noise with a PSD following a $f^\alpha$ power law but it keeps only a randomly selected sequence of $N$ consecutive data. The output of this software is a file containing this $N$-data sequence. This software, ``bruiteur'', is available on request by sending an email to the corresponding author.

\begin{figure}
\begin{tabular}{rl}
\includegraphics[scale=0.57]{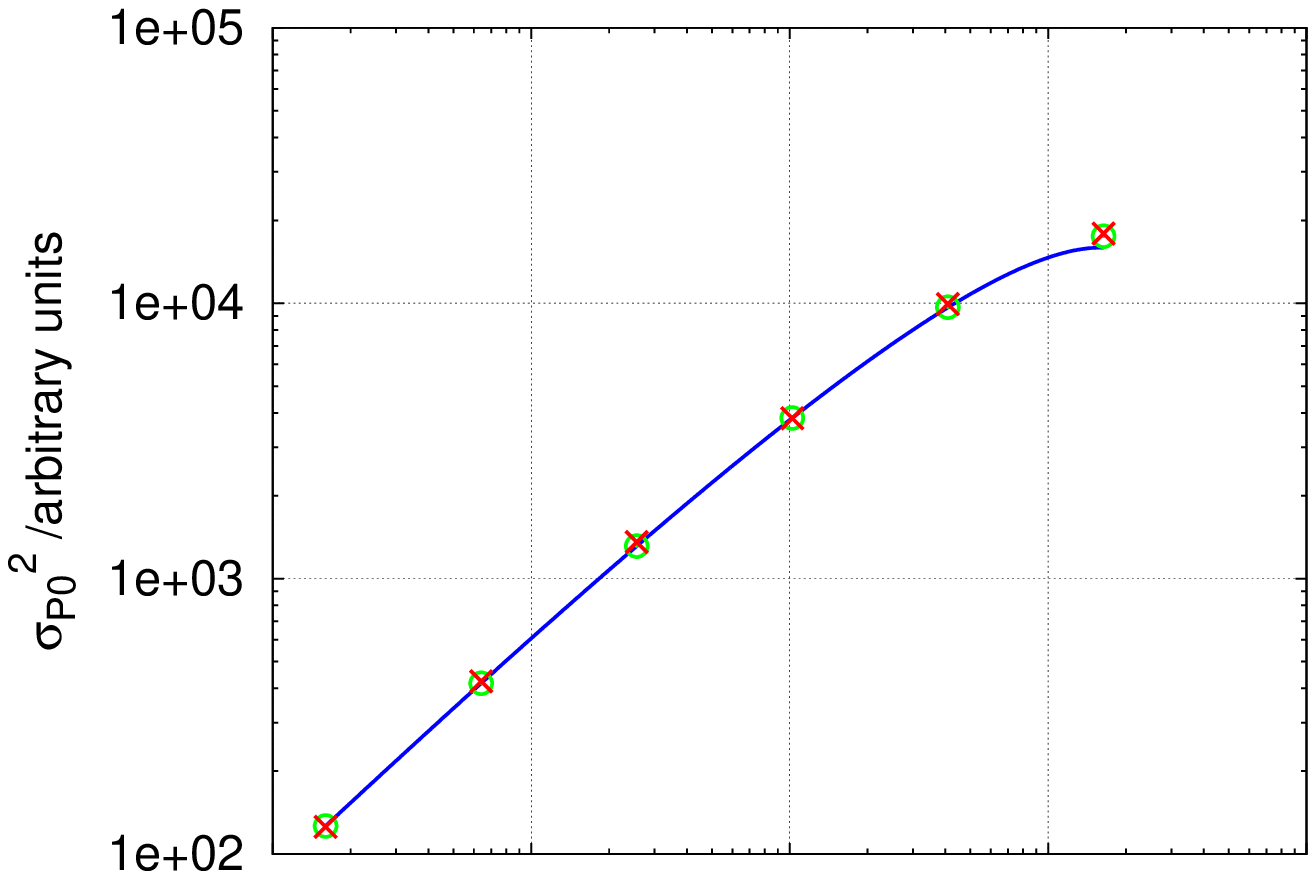} & \includegraphics[scale=0.57]{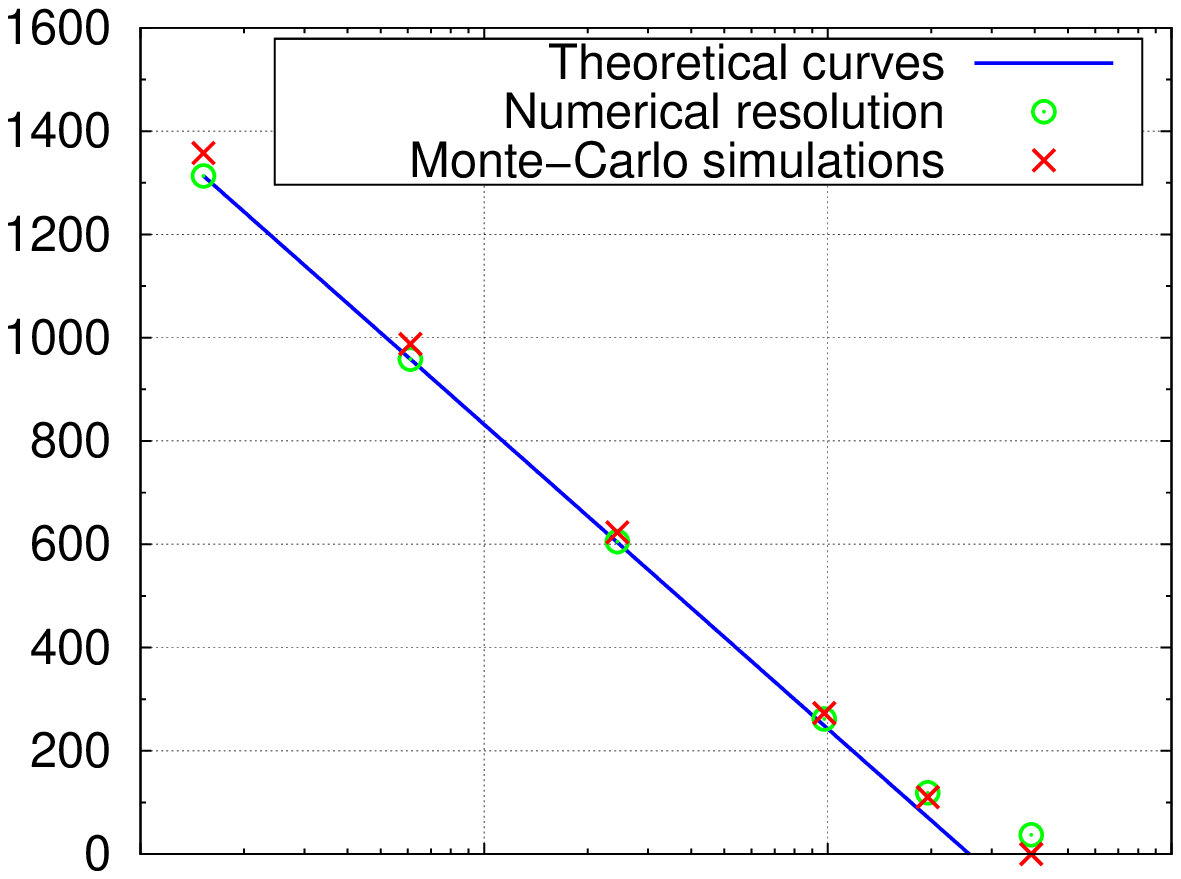}\\
\includegraphics[scale=0.57]{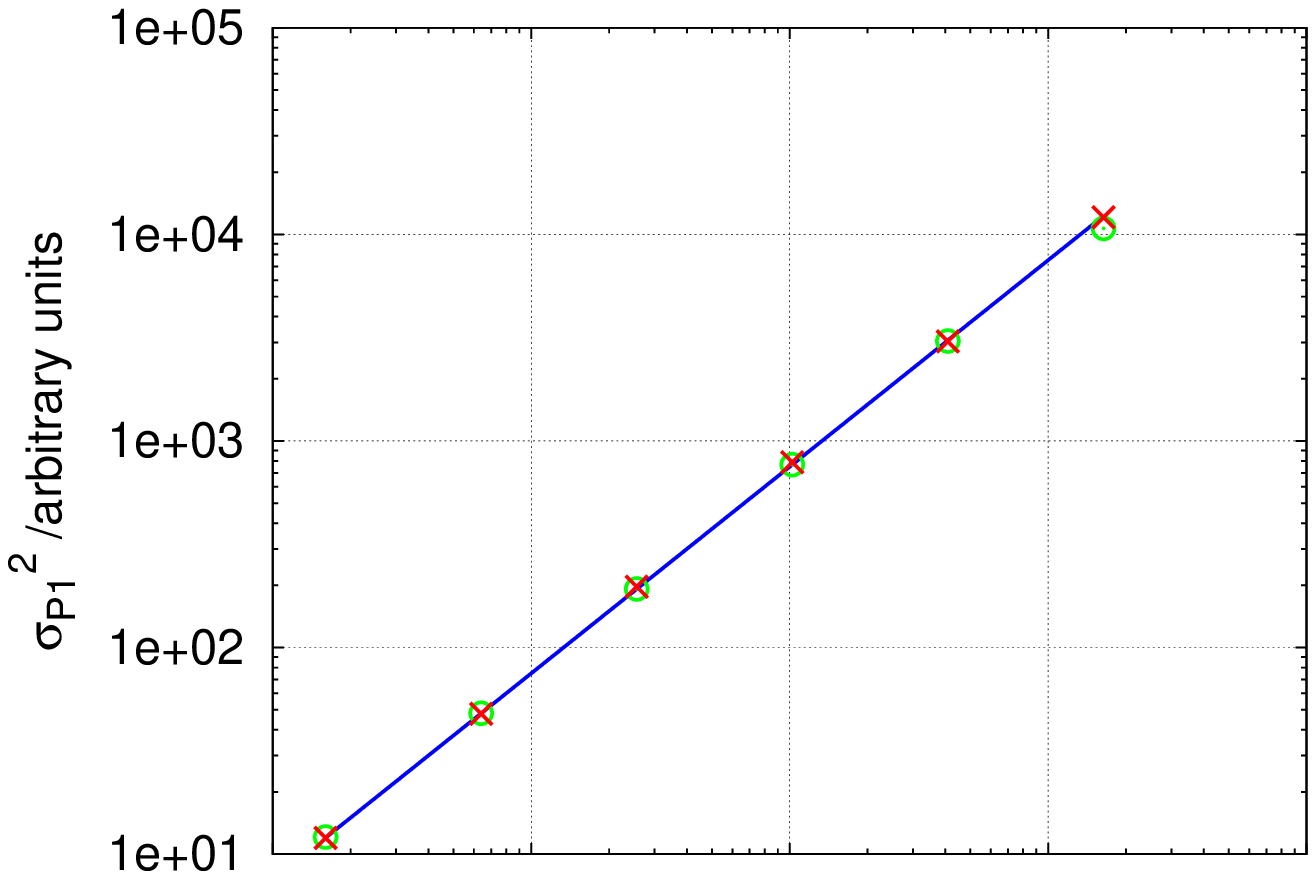} & \includegraphics[scale=0.57]{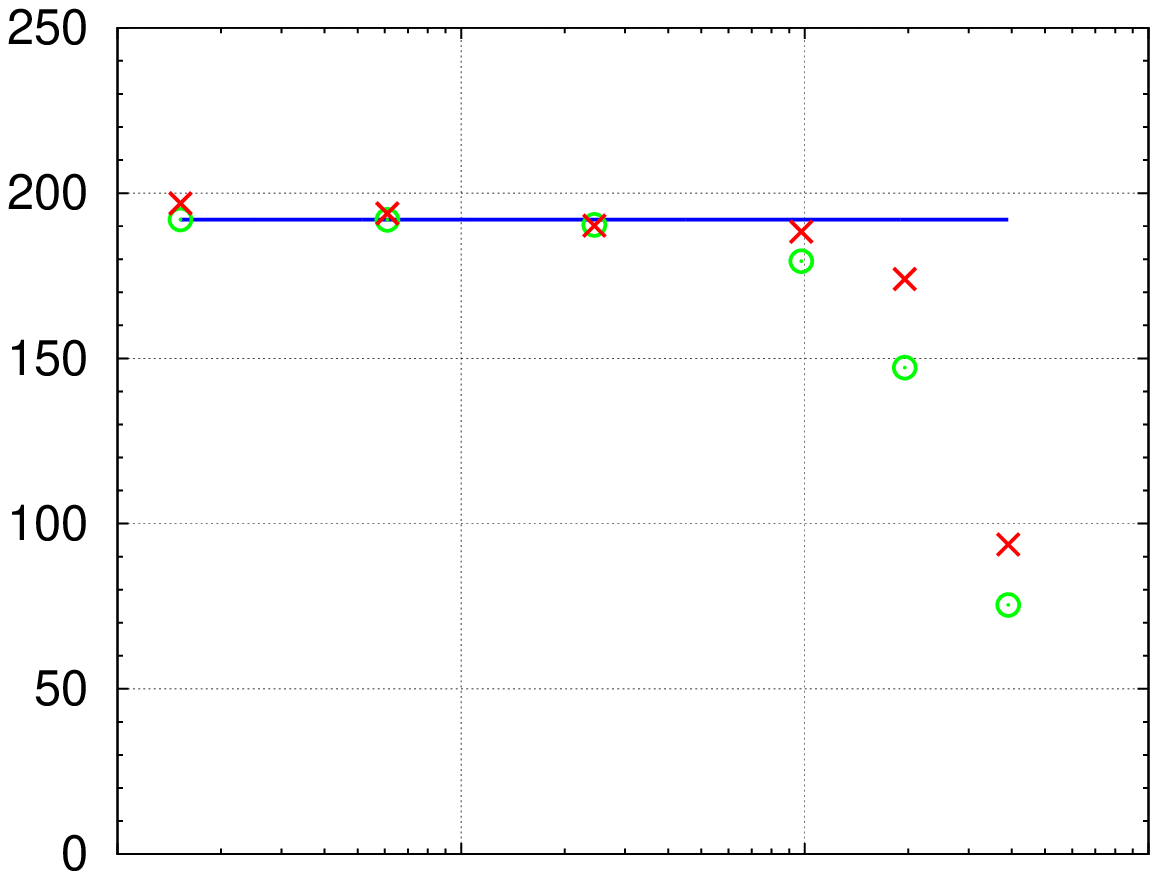}\\
\includegraphics[scale=0.57]{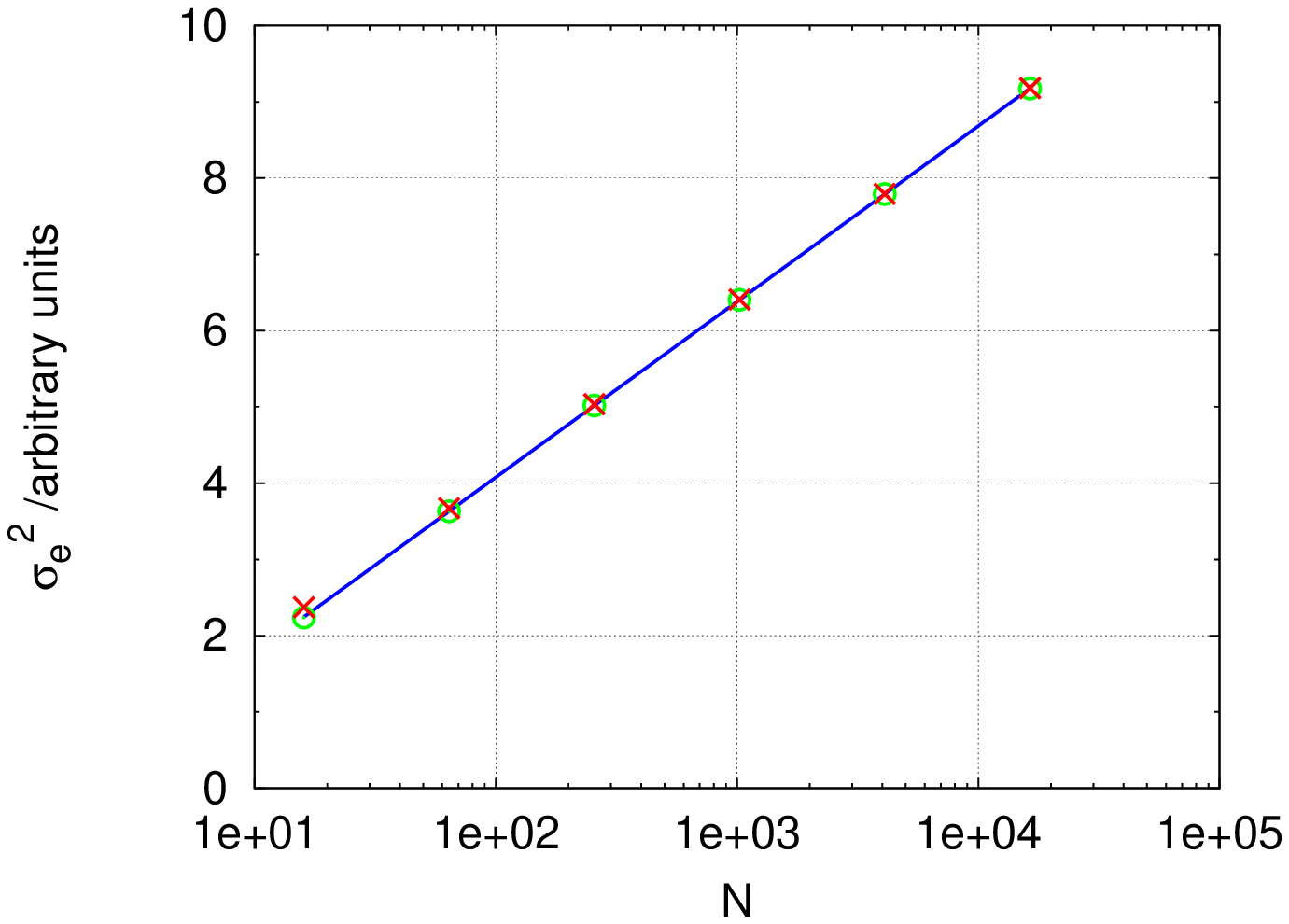} & \includegraphics[scale=0.57]{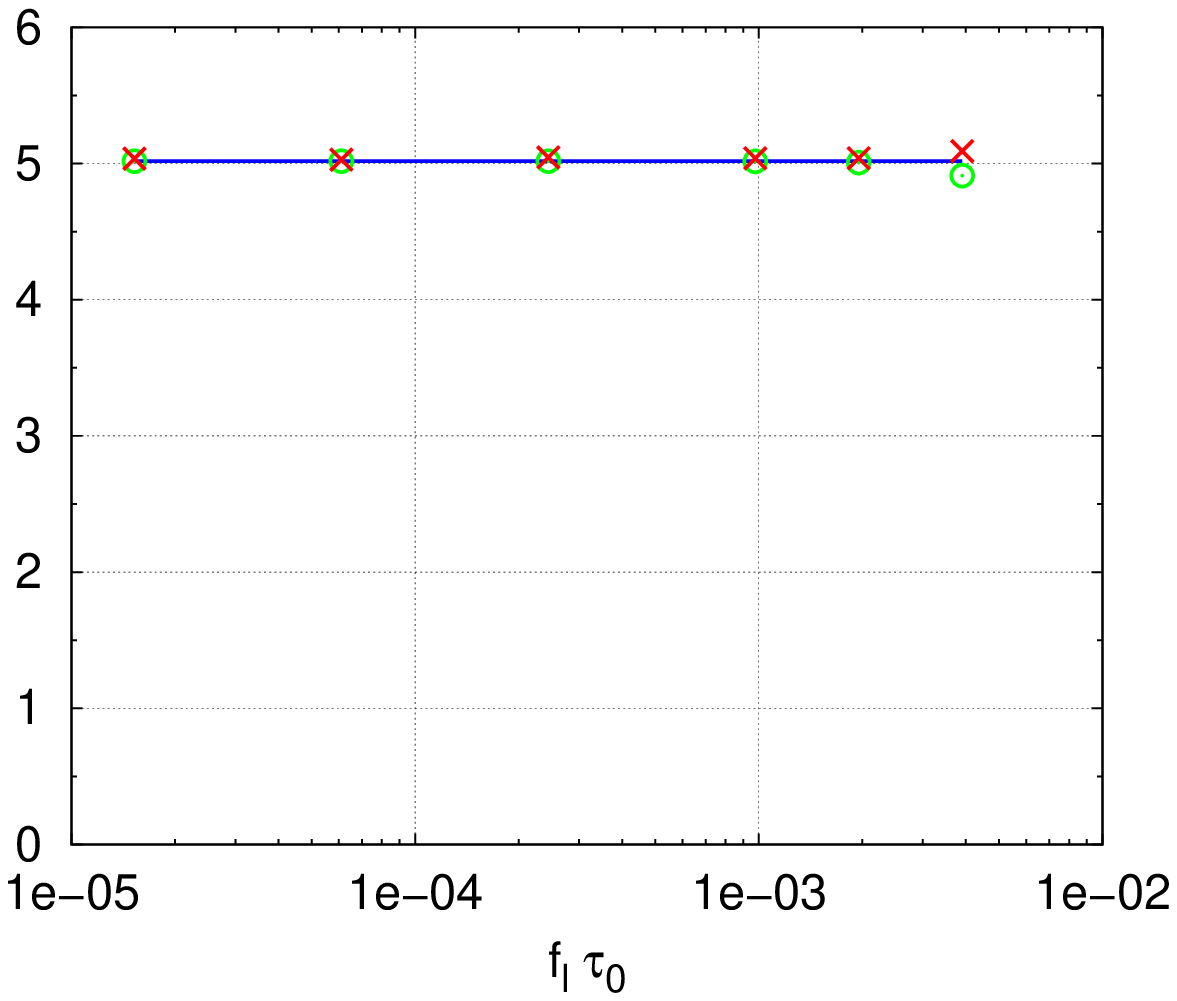}
\end{tabular}
\caption{Behavior of the variance of the $P_0$ parameter (above),  of the $P_1$ parameter (middle) and of the residuals (below) versus the number of data $N$ (left) and the low cut-off frequency $f_l$ (right). On the left side, $f_l=1/(65\ 536\ \tau_0)$. On the right side, $N=256$. The blue curves are plotted according to our theoretical results expressed in (\ref{eq:var_P0_fm1}), (\ref{eq:var_P1_fm1}) and (\ref{eq:var_res_fm1}). The green circles were obtained by numerical resolution. Each red cross is the average of the variance estimates obtained for 10\ 000 realizations of the same process. \label{fig:P01rsvsNfl}}
\end{figure}

Thus, thanks to this software, we used the same protocols than with the numerical computation: we first generated noise sequences for $N \in \{16, 64, 256, 1024, 4096, 16384\}$ data and $fl=1/(65536 \tau_0)$. Then, for $N=256$ data, we generated noise sequences for $\tau_0/f_l \in \{256, 512, 1024, 4096, 16384, 65536\}$. For each of these couple of $(N,f_l)$-values, $10\ 000$ sequences were generated, the $P_0$ and $P_1$ coefficients were calculated for each sequence, and the variance of these coefficients were calculated over the $10\ 000$ sequences. The results are plotted in figure \ref{fig:P01rsvsNfl} (red crosses) and compared to the theoretical values (blue curves) given by relationships (\ref{eq:var_P0_fm1}), (\ref{eq:var_P1_fm1}) and (\ref{eq:var_res_fm1}) and to the numerical computations (green circles).

	\subsection{Discussion\label{sec:discuss}}
		\subsubsection{Variation versus $N$.}
The left-hand side of figure \ref{fig:P01rsvsNfl} shows a very good agreement between the theoretical curves obtained from our theoretical relationships, the numerical computations and the Monte-Carlo simulations, proving that the dependence of $\sigma_{P0}^2$, $\sigma_{P1}^2$ and $\sigma_{e}^2$ versus $N$ are correctly modeled by (\ref{eq:var_P0_fm1}), (\ref{eq:var_P1_fm1}) and (\ref{eq:var_res_fm1}).

A slight discrepancy may be noticed for $N=16$ and $N=16384$. The latter is due to the proximity between the length of the sequence ($16384 \tau_0$) and the inverse of the low cut-off frequency ($65536 \tau_0$) and will be addressed in the next section (see {\S} \ref{sec:vsfl}).

On the other hand, the difference for $N=16$ is due to the small number of samples which should prohibit the neglecting of $N^{k-1}$ with respect to $N^k$, as we did for instance in (\ref{eq:approx_bigN}). However, table \ref{tab:vsN} shows that the discrepancy remains below the 10 \% level which is perfectly satisfactory for an uncertainty assessment. 

Therefore, the approximations of $\sigma_{P0}^2$, $\sigma_{P1}^2$ and $\sigma_{e}^2$ by (\ref{eq:var_P0_fm1}), (\ref{eq:var_P1_fm1}) and (\ref{eq:var_res_fm1}) may be considered valid for values of $N$ as small as 16.

\begin{table}
\caption{Comparison of the theoretical values obtained according to (\ref{eq:var_P0_fm1}), (\ref{eq:var_P1_fm1}), (\ref{eq:var_res_fm1}) to the numerical computations and to the Monte-Carlo simulation values of $\sigma_{P0}^2$, $\sigma_{P1}^2$ and $\sigma_e^2$ for $N=256$ and $f_l=1/(65536\tau_0)$. The percentages in brackets indicate the deviations of the theoretical values from the references (numerical and simulated). \label{tab:vsN}}
\begin{indented}
	\item \begin{tabular}{cccccc}
\br
$N=16$ & Theoretical & \multicolumn{2}{c}{Numerical} & \multicolumn{2}{c}{Simulation} \\
$f_l=1/(65536\tau_0)$ &&&&&\\
\mr
$\sigma_{P0}^2$ & 126.4 &  126.5 & (-0.08 \%) & 125.6 & (+0.7 \%)\\
$\sigma_{P1}^2$ & 12.00 & 12.08 & (-0.7 \%) & 11.96 & (+1 \%)\\
$\sigma_e^2$ & 2.244 & 2.237 & (+0.3 \%) & 2.373 & (-6 \%)\\
\br
\end{tabular}
\end{indented}
\end{table}

		\subsubsection{Variation versus $f_l$.\label{sec:vsfl}}
The agreement between the theoretical curves obtained from our theoretical relationships, the numerical computations and the Monte-Carlo simulations is less convincing when we observe the dependence versus the low cut-off frequency $f_l$ (see right-hand side of figure \ref{fig:P01rsvsNfl}). In particular, for the last three values of $f_l$, obtained for $1/(4T)$, $1/(2T)$ and $1/T$ (where $T=N\tau_0$ is the total duration of the sequence), we observe an increasing gap between the curves and the points.

This should not surprise us concerning $\sigma_{P0}^2$ since we know that $\sigma_{P0}^2=0$ if $f_l=1/T$. The approximation (\ref{eq:var_P0_fm1}) was clearly designed for being valid only if $f_l\ll 1/T$ and provides even a negative value for $\sigma_{P0}^2$ if $f_l=1/T$!

On the other hand, the discrepancy concerning $\sigma_{P1}^2$ is more puzzling since it was expected to not depend on $f_l$, according to the moment condition (see {\S} \ref{sec:moment_condition}). Furthermore, this seems to contradict our conception of the low cut-off frequency which was defined in {\S} \ref{sec:fl} as ``the inverse of the duration over which we subtract the arithmetic mean''. How the subtraction of a constant value could impact the variance of the linear drift coefficient?

This apparent paradox is removed if we distinguish a sequence whose low cut-off frequency is ``truly'' equal to $1/T$ from a sequence whose cut-off frequency was ``artificially'' set to $1/T$. In the first case, which happens in the noise simulator we used as well as in the numerical computation, the PSD of the sequence is strictly conform to the model expressed in (\ref{eq:PSD}). 

\begin{figure}
\centering{\includegraphics[width=7.5cm]{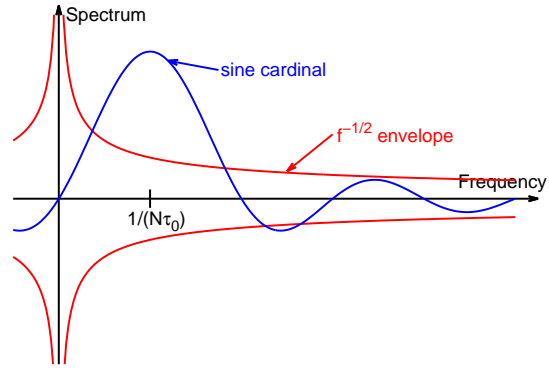}}
\caption{Influence of the very low frequency amplitudes of the spectrum on the first lobe of the Fourier transform of the window for $f=1/(N\tau_0)$.\label{fig:PSD_bavage}}
\end{figure}

In the other case, we are faced to a flicker sequence whose low cut-off frequency is probably very low, we may even consider that $f_l\to 0$, and we remove its arithmetic mean. But this sequence of duration $T$ has been extracted from a flicker noise sequence of far longer duration $\Theta$ and we may even consider that $\Theta\to \infty$. Such an extraction may be modeled in the direct domain by the multiplication of a $\Theta$-sequence by a $T$-window. In the Fourier domain, this operation amounts to perform a convolution product between the spectrum of the $\Theta$-sequence by the the Fourier transform of the $T$-window, i.e. by a narrow sine cardinal. Let us consider the first frequency sample greater than zero (we don't care about the amplitude of the null frequency since it will be set to zero), i.e. $f=1/(N\tau_0)$:
\begin{equation}
\tilde{d}\left(\frac{1}{N\tau_0}\right)=\int_{-\infty}^{+\infty}{\tilde{D}(f)\frac{\sin\left[\pi \tau_0 \left(f-\frac{1}{N\tau_0}\right)\right]}{\pi \left(f-\frac{1}{N\tau_0}\right)} \mathrm{d}f}
\end{equation}
where $\tilde{d}(f)$ and $\tilde{D}(f)$ are the Fourier transform of, respectively, the $T$-sub-sequence  and of the $\Theta$-sequence. Since the very low frequency amplitudes of $\tilde{D}(f)$ may be very high, their impact on the first lobe of the cardinal sine may be predominating (see figure \ref{fig:PSD_bavage}). Thus, the amplitudes of $\tilde{d}(f)$ for the frequency $1/(N\tau_0)$ and its first multiples may be ``polluted'' by the very low frequency amplitudes of  $\tilde{D}(f)$ and its PSD may significantly depart from the $1/f$ theoretical model. In that sense, such a sequence is not a ``true'' $1/f$ noise. But it is a ``truly realistic'' flicker noise because we will never encounter a ``true'' $1/f$ noise with a low cut-off frequency exactly equal to the inverse of the sequence duration!

To summarize, removing the arithmetic mean of a flicker sequence is equivalent to setting its low cut-off frequency to the inverse of the sequence duration but at the price of a slight deviation of its spectrum from a perfect flicker spectrum due to this pollution effect by the very low frequencies. 

Nevertheless, table \ref{tab:vsfl} shows that the behavior of $\sigma_{P0}^2$, $\sigma_{P1}^2$ and $\sigma_e^2$ is pretty well fitted by the approximations (\ref{eq:var_P0_fm1}), (\ref{eq:var_P1_fm1}) and (\ref{eq:var_res_fm1}) for $f_l \leq 1/(4N\tau_0)$ since the differences remains below 10 \%.

\begin{table}
\caption{Comparison of the theoretical values obtained according to (\ref{eq:var_P0_fm1}), (\ref{eq:var_P1_fm1}), (\ref{eq:var_res_fm1}) to the numerical computations and to the Monte-Carlo simulation values of $\sigma_{P0}^2$, $\sigma_{P1}^2$ and $\sigma_e^2$ for $N=256$ and $f_l=1/(1024\tau_0)$. The percentages in brackets indicate the deviations of the theoretical values from the references (numerical and simulated).\label{tab:vsfl}}
\begin{indented}
	\item \begin{tabular}{cccccc}
\br
$N=256$ & Theoretical & \multicolumn{2}{c}{Numerical} & \multicolumn{2}{c}{Simulation} \\
$f_l=1/(1024\tau_0)$ &&&&&\\
\mr
$\sigma_{P0}^2$ & 248.6 & 261.4 & (-5 \%) & 273.0 & (-9 \%)\\
$\sigma_{P1}^2$ & 192.0 & 179.4 & (+7 \%) & 188.3 & (+2 \%)\\
$\sigma_e^2$ & 5.017 & 5.016 & (+0.02 \%) & 5.039 & (-0.4 \%)\\
\br
\end{tabular}
\end{indented}
\end{table}

	\subsection{Comparison with Generalized Least Squares\label{sec:cmp_opt}}
\begin{figure}
\begin{tabular}{rl}
\includegraphics[scale=0.57]{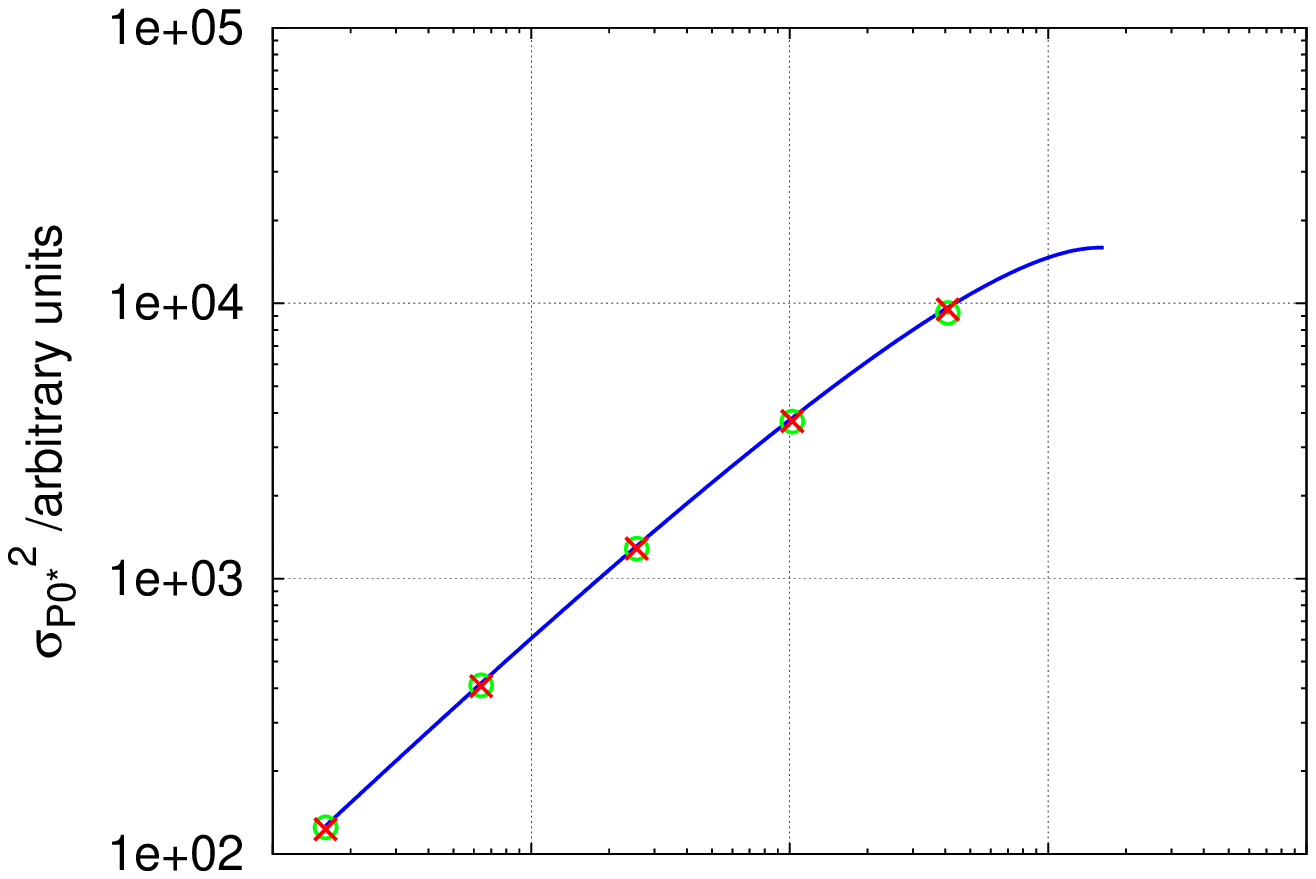} & \includegraphics[scale=0.57]{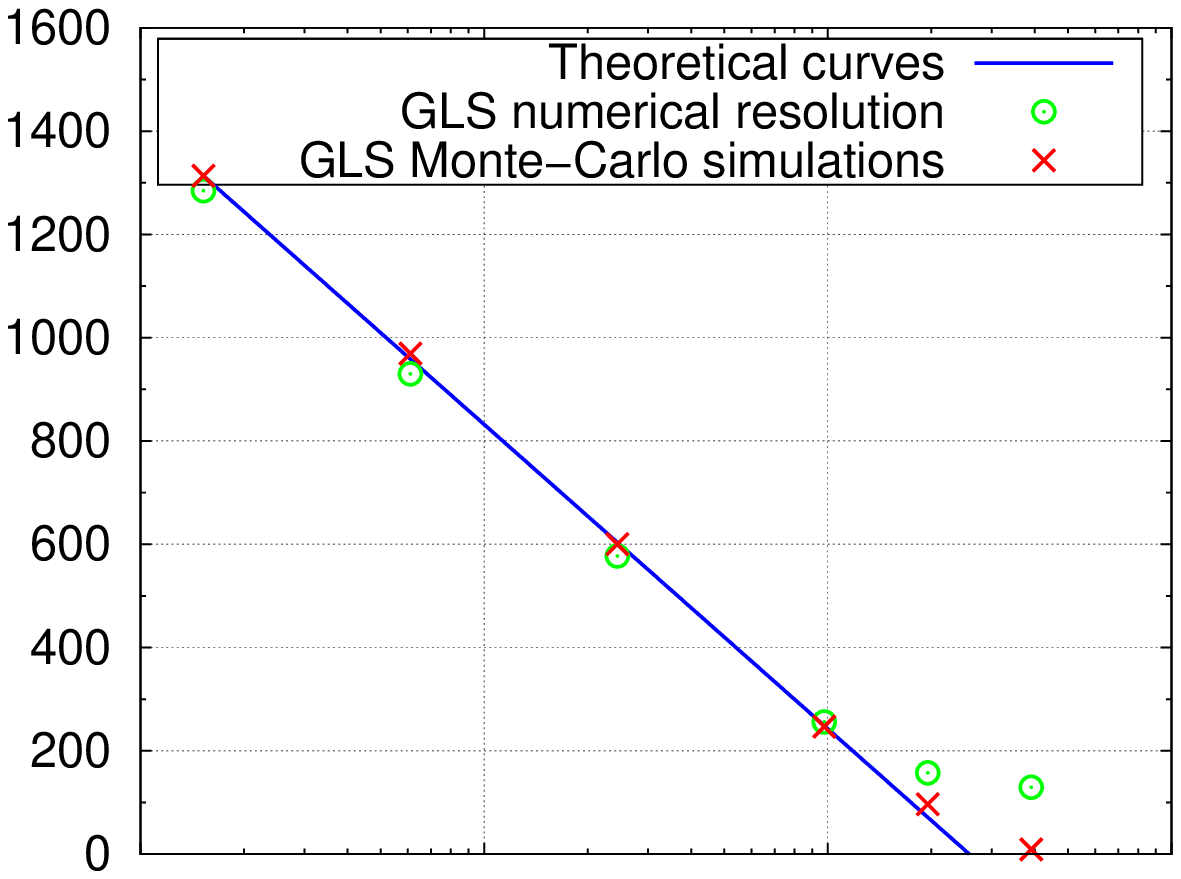}\\
\includegraphics[scale=0.57]{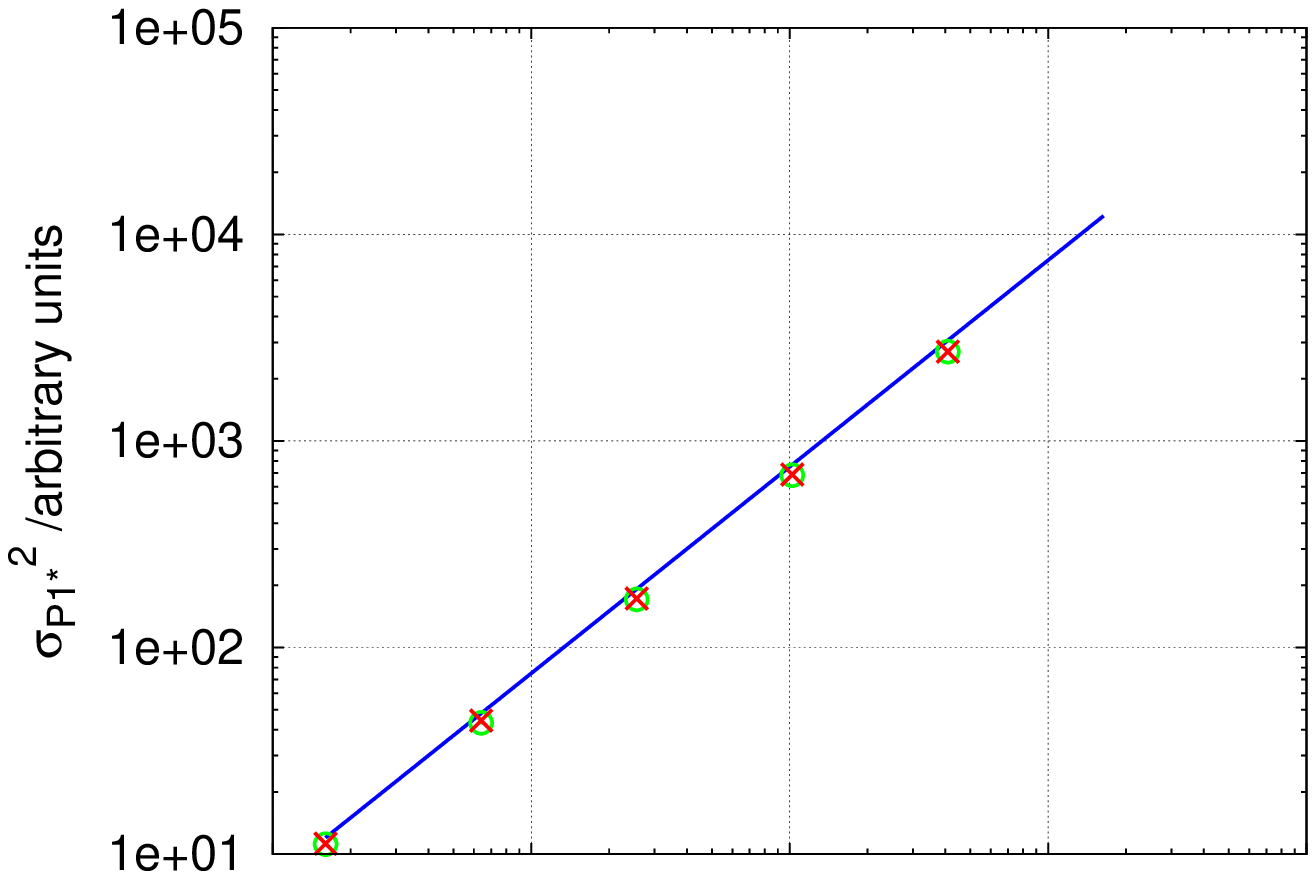} & \includegraphics[scale=0.57]{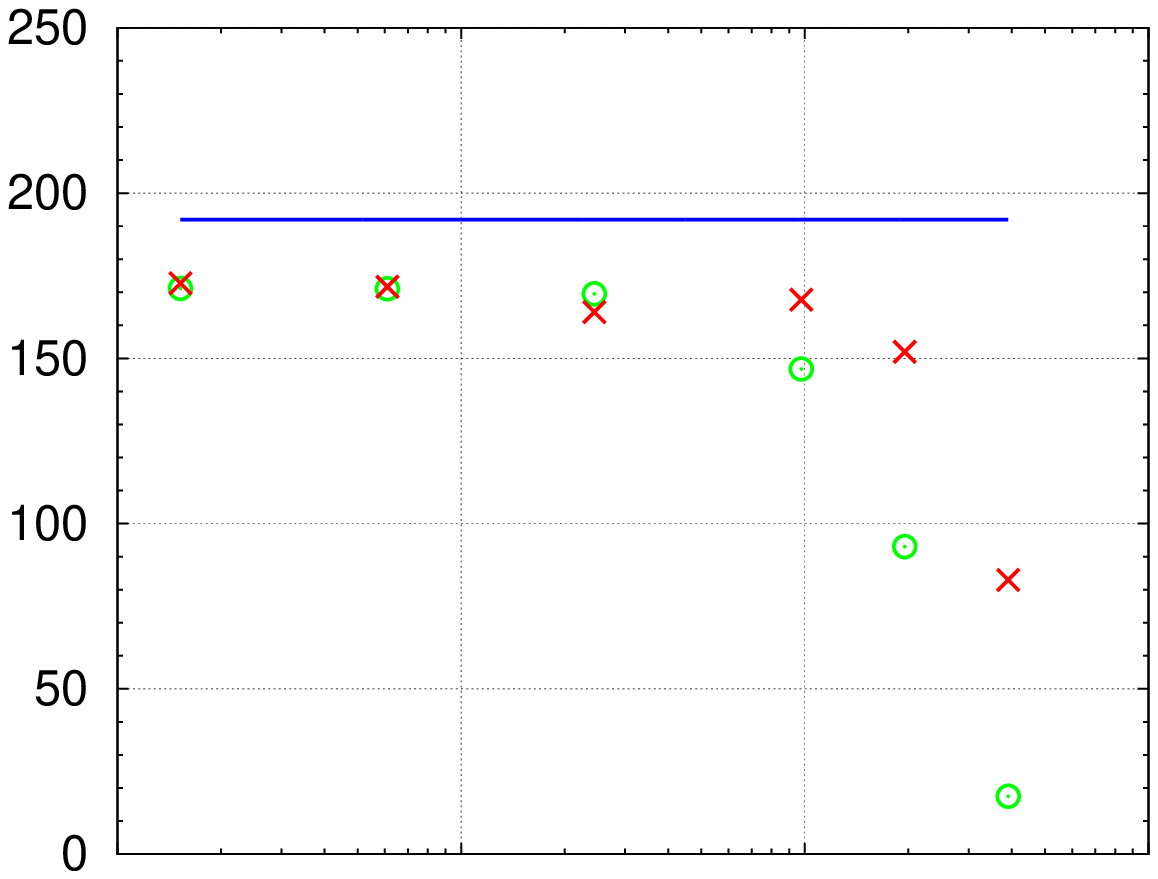}\\
\includegraphics[scale=0.57]{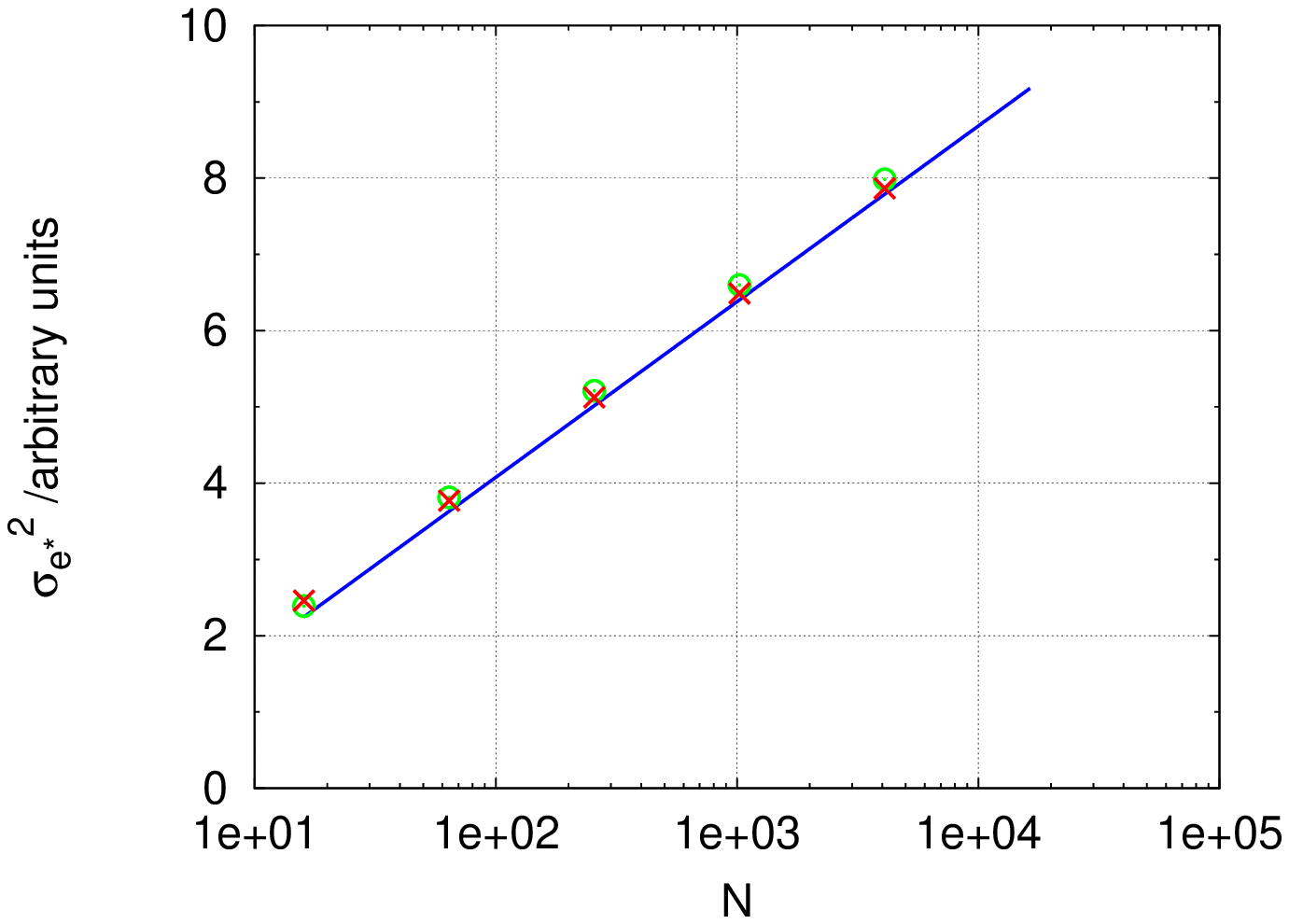} & \includegraphics[scale=0.57]{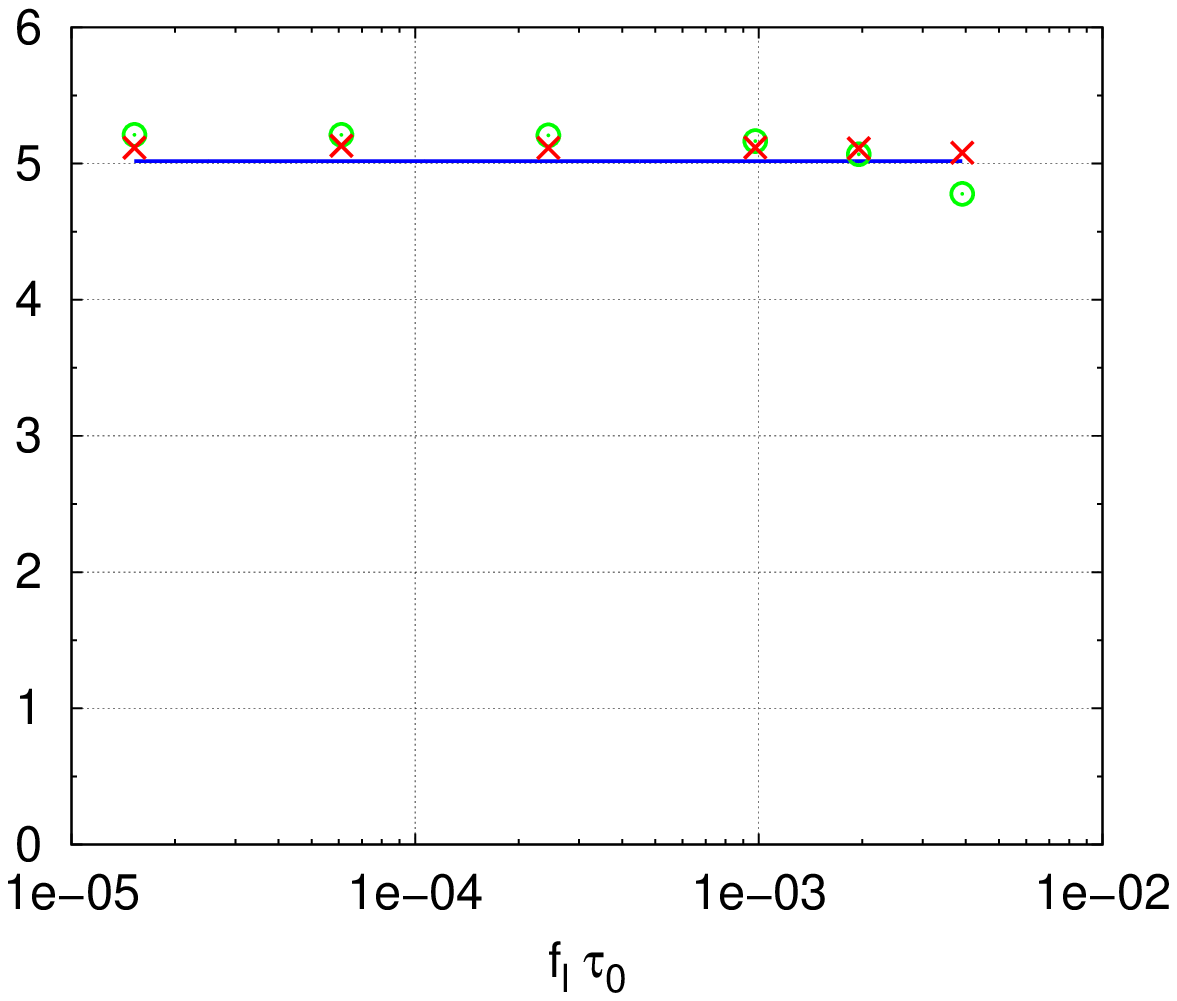}
\end{tabular}
\caption{Behavior of the variance of the $P_0^\star$ parameter (above),  of the $P_1\star$ parameter (middle) and of the residuals (below) versus the number of data $N$ (left) and the low cut-off frequency $f_l$ (right). On the left side, $f_l=1/(65\ 536\ \tau_0)$. On the right side, $N=256$. The blue curves are plotted according to our theoretical results expressed in (\ref{eq:var_P0_fm1}), (\ref{eq:var_P1_fm1}) and (\ref{eq:var_res_fm1}). The green circles were obtained by GLS numerical resolution. Each red cross is the average of the variance GLS estimates obtained for 10\ 000 realizations of the same process. \label{fig:P01rsvsNflGLS}}
\end{figure}

In order to compare the GLS estimates (\ref{eq:opt_GLS}), their variances (\ref{eq:varGLS}) and the variance of the residuals (\ref{eq:resGLS}) to the approximation given by the Chebychev Least Squares in (\ref{eq:estPk}), (\ref{eq:var_P0_fm1}), (\ref{eq:var_P1_fm1}) and (\ref{eq:var_res_fm1}), we used the same two types of checking as previously:
\begin{itemize}
	\item a comparison with numerical computations of (\ref{eq:varGLS}) and (\ref{eq:resGLS})
	\item a comparison with Monte-Carlo simulations obtained with a noise simulator (``bruiteur'').
\end{itemize} 
In all cases, we considered flicker noises with a unity level and we used the same two protocols as previously:
\begin{itemize}
	\item varying $N$ with $f_l$ constant; we chose $N \in \{16, 64, 256, 1024, 4096\}$ data and $fl=1/(65536 \tau_0)$
	\item varying $f_l$ with $N$ constant; we chose $\tau_0/f_l \in \{256, 512, 1024, 4096, 16384, 65536\}$ and $N=256$.
\end{itemize}
The results are plotted in figure \ref{fig:P01rsvsNflGLS} (green circles for the numerical computations and red crosses for the Monte-Carlo simulations) and compared to the theoretical values (blue curves) given by relationships (\ref{eq:var_P0_fm1}), (\ref{eq:var_P1_fm1}) and (\ref{eq:var_res_fm1}).

Despite the effects due to the low cut-off frequency which were already commented in {\S} \ref{sec:discuss}, figure \ref{fig:P01rsvsNflGLS} shows that the variance of the parameters $P_0^\star$ and $P_1^\star$ are slightly lower than the theoretical curves, as expected since the GLS is optimal. On the other hand, the variance of the residuals is slightly higher than the theoretical curve of $\sigma_e^2$ which is consistent with (\ref{eq:var_res}).

However, these differences are very small: looking at the left-hand side of figure \ref{fig:P01rsvsNflGLS} (variation versus $N$), the averaged overestimations of the theoretical curves are respectively +3 \%, +11 \% and -4 \% (underestimation in this case) for $\sigma_{P0\star}^2$, $\sigma_{P1\star}^2$ and $\sigma_{e\star}^2$. This result is confirmed by table \ref{tab:vsNGLS}.

\begin{table}
\caption{Comparison of the theoretical values obtained according to (\ref{eq:var_P0_fm1}), (\ref{eq:var_P1_fm1}), (\ref{eq:var_res_fm1}) to the GLS numerical computations and to the GLS Monte-Carlo simulation values of $\sigma_{P0\star}^2$, $\sigma_{P1\star}^2$ and $\sigma_{e\star}^2$ for $N=256$ and $f_l=1/(65536\tau_0)$. The percentages in brackets indicate the deviations of the theoretical values from the references (numerical and simulated). \label{tab:vsNGLS}}
\begin{indented}
	\item \begin{tabular}{cccccc}
\br
$N=16$ & Theoretical & \multicolumn{2}{c}{Numerical} & \multicolumn{2}{c}{Simulation} \\
$f_l=1/(65536\tau_0)$ &&&&&\\
\mr
$\sigma_{P0\star}^2$ & 126.4 &  125.0 & (+1.1 \%) & 123.1 & (+3 \%)\\
$\sigma_{P1\star}^2$ & 12.00 & 11.16 & (+8 \%) & 11.23 & (+7 \%)\\
$\sigma_{e\star}^2$ & 2.244 & 2.387 & (-6 \%) & 2.461 & (-9 \%)\\
\br
\end{tabular}
\end{indented}
\end{table}

This overestimation of the variances $\sigma_{P0\star}^2$ and $\sigma_{P1\star}^2$ by the theoretical curves is more visible on the right-hand side of figure \ref{fig:P01rsvsNflGLS} (variation versus $f_l$), particularly with the parameter $P_1^\star$ when $1/f_l$ tends toward $N\tau_0$ ($f_l > 10^{-3}\tau_0^{-1}$). This result is confirmed by table \ref{tab:vsflGLS} which exhibits a 30 \% overestimation if $\sigma_{P1\star}^2$ is approximated by (\ref{eq:var_P1_fm1}) with $f_l = 1/(4N\tau_0)$.

\begin{table}
\caption{Comparison of the theoretical values obtained according to (\ref{eq:var_P0_fm1}), (\ref{eq:var_P1_fm1}), (\ref{eq:var_res_fm1}) to the GLS numerical computations and to the GLS Monte-Carlo simulation values of $\sigma_{P0\star}^2$, $\sigma_{P1\star}^2$ and $\sigma_{e\star}^2$ for $N=256$ and $f_l=1/(1024\tau_0)$. The percentages in brackets indicate the deviations of the theoretical values from the references (numerical and simulated).\label{tab:vsflGLS}}
\begin{indented}
	\item \begin{tabular}{cccccc}
\br
$N=256$ & Theoretical & \multicolumn{2}{c}{Numerical} & \multicolumn{2}{c}{Simulation} \\
$f_l=1/(1024\tau_0)$ &&&&&\\
\mr
$\sigma_{P0\star}^2$ & 248.6 & 255.8 & (-3 \%) & 246.8 & (+0.8 \%)\\
$\sigma_{P1\star}^2$ & 192.0 & 146.8 & (+30 \%) & 167.7 & (+15 \%)\\
$\sigma_{e\star}^2$ & 5.017 & 5.166 & (-3 \%) & 5.118 & (-2 \%)\\
\br
\end{tabular}
\end{indented}
\end{table}


As a conclusion of this comparison with the GLS estimation, once the values for high $f_l$  excluded ($f_l\geq \frac{1}{4 N\tau_0}$), the difference between $\sigma_{P1\star}^2$ and the theoretical estimation given by (\ref{eq:var_P1_fm1}) remains below 20 \% (15 \% on average). This difference is no more than 3 \% concerning $\sigma_{P0\star}^2$ and (\ref{eq:var_P0_fm1}) as well as $\sigma_{e\star}^2$ and (\ref{eq:var_res_fm1}). Since an uncertainty is always roughly estimated, the theoretical estimation of the variances given by  (\ref{eq:var_P0_fm1}), (\ref{eq:var_P1_fm1}) and (\ref{eq:var_res_fm1})
may be considered as valid even for the GLS parameter and residual variances.

Therefore, the theoretical estimation given in this paper may be used for estimating the variance obtained by GLS, allowing thus to have a quick value without computing the trace of the $N\times N$ matrix $[C_\epsilon] - [\Phi][\Xi][\Phi]^T$. In particular, when $N$ is large (typically $N > 5000$), operations with such matrix is very time consuming or impossible (this is why the GLS estimation has not been performed for $N = 16384$ in figure \ref{fig:P01rsvsNflGLS}!).

As a consequence, taking into account the low gain and the computational complexity of GLS, the estimation of linear parameters using the Chebyshev least squares, i. e. (\ref{eq:estPk}),  comes as a powerful and very simple alternative although slightly suboptimal. Finally, this method is still valid when a measurement sequence contains a mixture of white noise and flicker, which is almost always the case. 

\section{Application to uncertainty domain estimation}
	\subsection{Confidence interval over the classical drift coefficients}
From (\ref{eq:P0P1toC0C1}) and knowing that the covariance between $P_0$ and $P_1$ is null, it is possible to calculate the variances of the classical drift coefficients $\sigma_{C0}^2$ and $\sigma_{C1}^2$ from the variances of the Chebyshev polynomial coefficients $\sigma_{P0}^2$ and $\sigma_{P1}^2$:
\begin{equation}
\left\{\begin{array}{rclcl}
\sigma_{C0}^2&=& \displaystyle \frac{1}{N} \sigma_{P0}^2 + \frac{3(N-1)}{N(N+1)} \sigma_{P1}^2 &\approx& \displaystyle \frac{1}{N} (\sigma_{P0}^2 + 3\sigma_{P1}^2)\\
\sigma_{C1}^2&=& \displaystyle \frac{12}{(N-1)N(N+1)\tau_0^2} \sigma_{P1}^2&\approx&\displaystyle \frac{12}{N^3\tau_0^2} \sigma_{P1}^2.
\end{array}\right.\label{eq:sPtosC}
\end{equation}

		\subsubsection{Expressions for $fl\ll 1/(N\tau_0)$.}
From (\ref{eq:var_P0_fm1}) and (\ref{eq:var_P1_fm1}), we obtain :
\begin{equation}
\left\{\begin{array}{l}
\sigma_{C0}^2 = \displaystyle \left[\frac{17}{4}-C-\ln(2\pi f_l N\tau_0)\right]k_{-1}\\
\sigma_{C1}^2 = \displaystyle \frac{9}{N^2\tau_0^2} k_{-1}.
\end{array}\right.
\end{equation}

As previously, we can express these results in terms of the variance of the residuals rather than in terms of the noise level $k_{-1}$. From (\ref{eq:var_res_fm1}), it comes:
\begin{equation}
\left\{\begin{array}{l}
\sigma_{C0}^2 = \displaystyle \frac{\frac{17}{4}-C-\ln(2\pi f_l N\tau_0)}{-\frac{9}{4}+C+\ln(2\pi f_h N\tau_0)} \sigma_e^2\\
\sigma_{C1}^2 = \displaystyle \frac{9}{\left[-\frac{9}{4}+C+\ln(2\pi f_h N\tau_0)\right]N^2\tau_0^2} \sigma_e^2.
\end{array}\right.
\end{equation}

Replacing the high cut-off frequency $f_h$ by the Nyquist frequency $1/(2\tau_0)$, we obtain the estimation of the variance of the classical parameters:
\begin{equation}
\left\{\begin{array}{l}
\sigma_{C0}^2 = \displaystyle \frac{\frac{17}{4}-C-\ln(2\pi f_l N\tau_0)}{-\frac{9}{4}+C+\ln(N\pi)} \sigma_e^2\\
\sigma_{C1}^2 = \displaystyle \frac{9}{\left[-\frac{9}{4}+C+\ln(N\pi)\right]N^2\tau_0^2} \sigma_e^2.
\end{array}\right.
\end{equation}

The 95 \% confidence interval over the estimates of the parameters $C_0$ and $C_1$ are then:
\begin{equation}
\left\{\begin{array}{rclcl}
\Delta C_0 &=& \displaystyle 2\sqrt{\frac{\frac{17}{4}-C-\ln(2\pi f_l N\tau_0)}{-\frac{9}{4}+C+\ln(N\pi)}} \sigma_e & \approx & \displaystyle 2\sqrt{\frac{1.385-\ln(f_l N\tau_0)}{-0.5281+\ln(N\pi)}}\sigma_e\\
\Delta C_1 &=& \displaystyle\frac{6}{N\tau_0\sqrt{-\frac{9}{4}+C+\ln(N\pi)}} \sigma_e & \approx & \displaystyle\frac{6} {N\tau_0\sqrt{-0.5281+\ln(N\pi)}}\sigma_e.
\end{array}\right.\label{eq:confint_small_fl}
\end{equation}

Thus, the uncertainty over $C_0$ is approximately constant and depends very slightly on the number of measurements. The uncertainty over $C_1$ decreases approximately as $1/N$ (whereas it decreases as $1/N^{3/2}$ for a white noise). Therefore, increasing $N\tau_0$ does not improve significantly the accuracy of the $C_0$ estimation but improves the accuracy of the $C_1$ estimation. It remains then useful to increase the length of the measurement sequence in a flicker context. 

On the other hand, increasing $N$ in a constant $T$-duration sequence improves only very slightly the accuracy of both parameter estimations.

	\subsubsection{Expression for $f_l=1/(N\tau_0)$.}
Let us consider now the case of $f_l=1/(N\tau_0)$, i.e. after removing the arithmetic mean. Remember that in this case, we implicitly set $P_0$ to 0 and then its variance is identically null. Therefore, from (\ref{eq:sPtosC}), $\sigma_{C0}^2=3 \sigma_{P1}^2/N$. Using this relationship and replacing $f_l$ by $1/(N\tau_0)$, we find:
\begin{equation}
\left\{\begin{array}{rcl}
\sigma_{C0}^2 &=& \displaystyle \frac{9}{-9+4C+4\ln(2\pi)} \sigma_e^2\\
\sigma_{C1}^2 &=& \displaystyle \frac{9}{\left[-\frac{9}{4}+C+\ln(2\pi)\right]N^2\tau_0^2} \sigma_e^2.
\end{array}\right.
\end{equation}

The 95 \% confidence intervals over the estimates of $C_0$ and $C_1$ are then:
\begin{equation}
\left\{\begin{array}{rcl}
\Delta C_0 & \approx & \displaystyle \frac{3 \sigma_e}{\sqrt{-0.5281+\ln(N)}}\\
\Delta C_1 & \approx & \displaystyle \frac{6 \sigma_e}{N\tau_0\sqrt{-0.5281+\ln(N)}}.
\end{array}\right.\label{eq:conf_int_big_fl}
\end{equation}

	\subsection{Confidence interval over the estimate $\hat{D}$}
From (\ref{eq:estPk}), we see that the $P_0$ parameter is obtained as:
\begin{equation}
P_0=\frac{1}{\sqrt{N}}\sum_{i=0}^{N-1} d_i.
\end{equation}
Since the estimate $\hat{D}$ is the arithmetic mean of the $\{d_i\}$ sequence, $\hat{D}=P_0/\sqrt{N}$. Therefore, the confidence interval over $\hat{D}$ is:
\begin{equation}
\Delta D=\frac{\Delta P_0}{\sqrt{N}}.\label{eq:eqP0D}
\end{equation} 

		\subsubsection{Expressions for $fl\ll 1/(N\tau_0)$.}
From (\ref{eq:varP0P1Rs}) and (\ref{eq:eqP0D}), we find:
\begin{equation}
\Delta D= 2 \sqrt{\frac{-0.4151-\ln\left(f_l N\tau_0\right)}{-2,112 + 4 \ln(N)}}\sigma_e.\label{eq:confintD_small_fl}
\end{equation}

What is the purpose of this relationship? Suppose that we have a long measurement sequence of duration $\Theta=M\tau_0$ and that we want to estimate the mean value of this sequence, $D_\Theta$ from the arithmetic mean of a subset of this sequence of duration $T=N\tau_0$, $\hat{D}_T$. In this case, we can use (\ref{eq:confintD_small_fl}) by replacing $f_l= 1/\Theta$. The arithmetic mean of the whole sequence $D_\Theta$ should be within the interval $[\hat{D}_T -\Delta D, \hat{D}_T+\Delta T]$ at 95 \% confidence.

		\subsubsection{Expressions for $fl = 1/(N\tau_0)$.}
This is of course the most interesting case. But remember that if $fl = 1/(N\tau_0)$, then $\Delta P_0=0$ as well as $\Delta D$. Strictly speaking, this is true. As we already explained in {\S} \ref{sec:fl}, it means that we consider that our estimate $\hat{D}$ \textbf{is} the true value over $T=N \tau_0$. But, in order to ensure the continuity of the measurements to adjacent $T$-sequences (e.g. for a $T=1$ day-averaged measurement, the continuity of the measurement from yesterday to tomorrow), we'd better consider what happened in the close past and what will happen in the close future by contextualizing our current measurement sequence among the previous and the next one, i.e. by considering a $\Theta$-duration at least equal to $3T$ (e.g. for a $T=1$ day-averaged measurement, $\Theta=3$ days). 

On the other hand, we saw that the determination accuracy of $C_0$ and $C_1$ is almost independent on the number of samples in a given $T$-duration sequence. Similarly, the variance of the determination of $D$ does not depend on $N$ since this variance has been set to 0 by removing the mean. However, it is obvious than the results of the arithmetic mean for different $N$ values and a fixed duration $T$ cannot be exactly the same. How could we handle these differences with a confidence interval?

Thus, taking into account a low cut-off frequency equal to 3 or 4 times $N\tau_0$ in (\ref{eq:confintD_small_fl}) will ensure that the confidence interval obtained over the current sequence will be compatible with the previous and the next estimates as well as with estimates obtained for different values of $N$:
\begin{equation}
\Delta D= 2 \sqrt{\frac{-0.4151-\ln\left(4\right)}{-2,112 + 4 \ln(N)}}\sigma_e\approx \frac{\sigma_e}{\sqrt{-0.5+\ln(N)}}.\label{eq:confintD_big_fl}
\end{equation}
The use of (\ref{eq:confintD_big_fl}) is then nothing but a recommendation and  
is not based on a rigorous foundation.

	\subsection{Application to real experimental measurements}
\begin{figure}
\centering{\includegraphics[width=10cm]{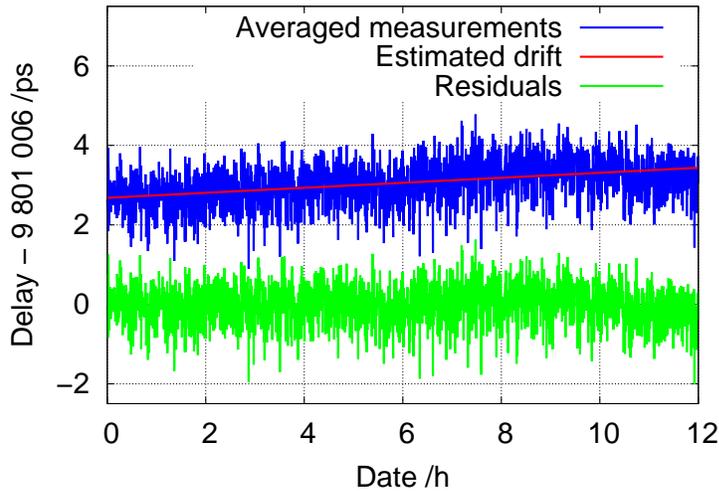}}
\caption{Processing of a sequence of delay measurements. A constant of 9\ 801\ 006 ps has been subtracted in order to compare the measurements (in blue) to the residuals (in green). The least square drift is figured in red.\label{fig:tr_li}}
\end{figure}

Figure \ref{fig:tr_li} presents an example of delay measurements affected by a flicker noise (see \cite{meyer2014} for the context of these measurements).

The questions are :
\begin{enumerate}
	\item Does this sequence exhibits a linear trend?
	\item If not, what is the confidence interval over the mean delay estimation?
\end{enumerate}

The parameter of this sequence are the following:
\begin{itemize}
	\item $N=2160$ data
	\item $\tau_0=20$ s
	\item $T=N\tau_0=12$ h
\end{itemize}

	\subsubsection{Rough results.}
\begin{itemize}
	\item Linear regression:
\begin{itemize}
	\item $C_0=9\ 801\ 008.68$ ps
	\item $C_1=1.75 \cdot 10^{-17}$ s/s $=1.51 $ ps/day
\end{itemize}
	\item Standard deviation of the residuals: $\sigma_e=0.51$ ps
	\item Arithmetic mean of the measurements: $\hat{D}=9\ 801\ 009.06$ ps.
\end{itemize}

	\subsubsection{Application of the confidence interval assessments.}
By using, respectively, the relationships (\ref{eq:conf_int_big_fl}) and (\ref{eq:confintD_big_fl}), we found:
\begin{itemize}
	\item $\Delta C_0=0.57$ ps
	\item $\Delta C_1=2.65 \cdot 10^{-17}$ s/s $=2.29$ ps/day
	\item $\Delta D=0.18$ ps.
\end{itemize}

	\subsubsection{Solution.}
The linear drift coefficient $C_1$ is within $1.51\pm 2.29$ ps/day at 95 \% confidence. Therefore, it is fully compatible with a null drift. We can then answer to the first question that no linear drift is detected in this sequence.

The confidence interval over the whole sequence is: $D= 9\ 801\ 009.06 \pm 0.18$ ps. This confidence interval should be compatible with a measurement of the same type performed over a 12 h-sequence immediately before or after this one.

	\subsubsection{Effect of decimation.}
In order to observe the impact of $N$ for a given duration, we decimated the number of samples by $3$ ($N=720$), $10$ ($N=216$), $30$ ($N=72$) and $108$ ($N=20$). We obtained the following results:
\begin{enumerate}
	\item $N=720$:
	\begin{itemize}
		\item $C_0= 9801008.73 \pm 0.61$ ps @ 95 \%
		\item $C_1=1.29 \pm 2.44$ ps/day @ 95 \%
		\item $D= 9801009.05 \pm 0.19$ ps @ 95 \%
	\end{itemize}
	\item $N=216$:
	\begin{itemize}
		\item $C_0=9801008.77 \pm 0.72$ ps @ 95 \%
		\item $C_1=1.31 \pm 2.86$ ps/day @ 95 \%
		\item $D=9801009.09 \pm 0.22$ ps @ 95 \%
	\end{itemize}
	\item $N=72$:
	\begin{itemize}
		\item $C_0=9801008.65 \pm 0.83$ ps @ 95 \%
		\item $C_1=1.79 \pm 3.29$ ps/day @ 95 \%
		\item $D=9801009.09 \pm 0.24$ ps @ 95 \%
	\end{itemize}
	\item $N=20$:
	\begin{itemize}
		\item $C_0=9801008.72 \pm 1.12$ ps @ 95 \%
		\item $C_1=1.44 \pm 4.48$ ps/day @ 95 \%
		\item $D=9801009.07 \pm 0.31$ ps @ 95 \%
	\end{itemize}
\end{enumerate} 
Then, the confidence intervals increase by a factor less than 2 whereas $N$ is divided by 108. We note also 
that $\hat{D}$ is extremely stable versus $N$: it varies only of 0.04 ps, far lower than the confidence interval calculated according to (\ref{eq:confintD_big_fl}). This confidence interval estimation is designed for continuity over different sequences and not for dealing with decimation.

\section{Conclusion}
After discussing the physical meaning of the low cut-off frequency which must be introduced for ensuring the convergence of the statistical parameters in a context of duration dependent noise, we defined a realistic model of power spectral density for a flicker noise. From this, we deduced the autocorrelation function of this type of noise and then a theoretical estimation of the variance of the linear drift parameters as well as of the arithmetic mean for a flicker noise. Then, we compared this method of drift estimation with the Generalized Least Square and concluded that our method is much easier to use although slightly suboptimal. Once the theoretical relationships for estimating the drift parameter variances were validated by Monte-Carlo simulations and numerical computations, we were able to establish rigorously confidence intervals over both drift coefficients and a recommendation for the confidence interval over the arithmetic mean. Finally, a complete example of processing of a real measurement sequence was given. 

Two issues remain open: how could we model the deviation of a spectrum with a very low cut-off frequency from a perfect flicker spectrum? How could we rigorously handle the variations of the arithmetic mean due to decimation? These questions are not fundamental from the metrological point of view but are important for a thorough understanding of the notion of low cut-off frequency. 

More generally, the approach followed in this paper could be used for assessing the confidence intervals of various statistical parameters with different types of strongly correlated noises (random walk, $f^{-3}$, $f^{-4}$, \ldots noises). For this purpose, the moment condition is very useful since it establishes a correspondence between convergence for low frequency noises and sensitivity to drifts. This could be very useful for time and frequency metrology, but also in many other domains.

On the other hand, as it was already mentioned in a previous paper \cite{vernotte2002}, one may use these results (or other ones for other types of noises) not to estimate confidence intervals over actual measurements but, at the opposite, to simulate different types of noise in a realistic manner, much more optimized than the simulator we used in this paper. For example, it would be far better to simulate a very low cut-off frequency by adding the appropriate drift than by computing a very long noise sequence and keeping a very small subset of it.

\section*{Acknowledgements}
This work was partially funded by the Laser MegaJoule program of the French Commissariat \`{a} l'\'{E}nergie Atomique. The time and frequency team of UTINAM is partially funded by the ANR Programme d'Investissement d'Avenir (PIA) under the Oscillator IMP project and First-TF network, and by funds from the R\'{e}gion Franche Comt\'{e} intended to support the PIA.

The authors wish to thank Vincent Drouet and Michel Prat, from the CEA, for their fruitful collaboration whose this paper is derived as well as Nicolas Gautherot and \'{E}ric Meyer, from UTINAM, for their valuable help and for their patience to perform very sensitive measurements. We are also grateful to Fran\c{c}ois Meyer for having encouraged us to publish this work.

\section*{References}
\bibliography{flicker_metrologia}

\newpage

\renewcommand{\theequation}{A-\arabic{equation}}
\setcounter{equation}{0}
\section*{APPENDIX: Variance of an estimate calculated in the frequency domain}
Let us consider a measurement sequence $x(t)$ sampled with a sampling step $\tau_0$.
In the same way as (\ref{eq:estpi}), let us denote $\hat{\theta}(t_0)$ the estimate obtained by the interpolating function $h(t)$ applied to $x(t)$ at the date $t_0$\footnote{If the causality must be taken into account, i.e. in the case of real-time processing, (\ref{eq:a1est}) may be rewritten as:
$$
\hat{\theta}(t_0) = \sum_{k=0}^{N-1}h\left[\left(k-N-1\right)\tau_0\right] x\left[t_0+\left(k-N-1\right)\tau_0\right].
$$
On the other hand, if causality does not matter, we could either write (\ref{eq:a1est}):
$$
\hat{\theta}(t_0) = \sum_{k=0}^{N-1}h(k\tau_0) x(t_0+k\tau_0).
$$}:
\begin{equation}
\hat{\theta}(t_0) = \sum_{k=0}^{N-1}h\left[\left(k-\frac{N-1}{2}\right)\tau_0\right] x_k = \sum_{k=0}^{N-1}h_k x_k, 
\label{eq:a1est}
\end{equation}
where $x_k=x\left[t_0+\left(k-\frac{N-1}{2}\right)\tau_0\right]$. Obviously, whatever the estimator $\theta$ is (e.g. arithmetic mean, constant drift coefficient, linear drift coefficient, \ldots), its estimates $\hat{\theta}(t_0)$ will depend on $t_0$ but neither its mathematical expectation nor its variance should vary: we assume that $\theta$ is stationary in this sense (even if it depends on the duration of the sequence $N\tau_0$). 

Let us define the 
 interpolating function $h(t)$ as:
\begin{equation}
h(t)=\sum_{k=0}^{N-1}h_k \delta\left[t+\left(k-\frac{N-1}{2}\right)\tau_0\right].
\end{equation}

The estimate $\hat{\theta}(t_0)$ may be rewritten as a convolution product:
\begin{equation}
\hat{\theta}(t_0) = \int_{-\infty}^{+\infty} h(t) x(t-t_0) \mathrm{d}t = \left[h(t)\ast\overline{x(-t)}\right]_{(t_0)}
\label{eq:a1cont}
\end{equation}
where the over line denotes a conjugate complex.

If $\theta$ is a centered Gaussian random variable, its variance is given by\footnote{In a previous paper (see Appendix 1 of \cite{vernotte2001}), we established the equality:
$$
\mathrm{E}\left(\theta^2\right) = \mathrm{E}\left\{\left[\int_{-\infty}^{+\infty} h(t) x(t) \mathrm{d}t\right]^2\right\}=\int_{0}^{+\infty} \left|H(f)\right|^2 S_x(f) \mathrm{d}f.\label{eq:hpp}
$$
However, we have found recently that the demonstration was flawed. The result remains correct under certain conditions. The correct demonstration and the conditions where it is true are developed in this Appendix. }:
\begin{equation}
\mathrm{E}\left(\theta^2\right) = \mathrm{E}\left\{\left|\left[h(t)\ast\overline{x(-t)}\right]_{(t_0)}\right|^2\right\}.
\label{eq:a1var}
\end{equation} 

Assuming the mathematical expectation as both an ensemble average and a time average, (\ref{eq:a1var}) becomes:
\begin{equation}
\mathrm{E}\left(\theta^2\right) = \left<\lim_{T\to\infty}\frac{1}{T}\int_{-T/2}^{+T/2}\left|\left[h(t)\ast\overline{x(-t)}\right]_{(t_0)}\right|^2 \mathrm{d}t_0\right>.\label{eq:a1conv}
\end{equation}

Let us define the function $G(f,T)$ as the ``\textit{windowed Fourier transform}'' of a function $g(t)$:
\begin{equation}
G(f,T)=\int_{-T/2}^{+T/2}g(t) e^{-j2\pi ft}\mathrm{d} t,\label{eq:limFT}
\end{equation}

We can apply the Parseval-Plancherel theorem over (\ref{eq:a1conv}):
\begin{eqnarray}
\mathrm{E}\left(\theta^2\right) &=& \left<\lim_{T\to\infty}\frac{1}{T}\int_{-\infty}^{+\infty}\left|H(f)\cdot\overline{X(f)}\right|^2 \mathrm{d}f\right>\nonumber\\
&=&\left<\lim_{T\to\infty}\frac{1}{T}\int_{-\infty}^{+\infty}\left|H(f)\right|^2\left|X(f)\right|^2 \mathrm{d}f\right>
\label{eq:TFif}
\end{eqnarray}
where $H(f)$ and $X(f)$ are respectively the windowed Fourier transform of $h(t)$ and $x(t)$ as defined in (\ref{eq:limFT}).

Since, by definition, the two-sided PSD of $x(t)$ is:
\begin{equation}
S_x^{TS}(f)=\left<\lim_{T\to\infty}\frac{1}{T}\left|X(f)\right|^2\right>,\label{eq:defPSD}
\end{equation}
we can rewrite (\ref{eq:TFif}) as:
\begin{equation}
\mathrm{E}\left(\theta^2\right)=\int_{-\infty}^{+\infty}\left|H(f)\right|^2S_x^{TS}(f)\mathrm{d}f=\int_0^{+\infty}\left|H(f)\right|^2S_x^{OS}(f)\mathrm{d}f\label{eq:eqT-F}
\end{equation}
where $S_x^{TS}(f)$ and $S_x^{OS}(f)$ are respectively the``Two-Sided'' and the ``One-Sided'' PSD, see (\ref{eq:STS}).

To conclude, (\ref{eq:est_f}) is true if the the signal is stationary, meaning that the Fourier transform of its variance in the time domain, i.e. the autocorrelation of $X(f)$,  is a Dirac distribution in the frequency domain. It means that all expectations may be obtained by averaging many realizations of a sequence of independent samples in the frequency domain, with amplitude expectations corresponding to the considered noise (e.g. $1/f$). However, a slight correlation of the first frequency samples may occur due to the windowing effect as explained in {\S} \ref{sec:vsfl}. 

\end{document}